\documentclass[lettersize,journal]{IEEEtran}
\usepackage{amsmath,amsfonts}
\usepackage{algorithmic}
\usepackage{algorithm}
\usepackage{array}
\usepackage[caption=false,font=normalsize,labelfont=sf,textfont=sf]{subfig}
\usepackage{textcomp}
\usepackage{stfloats}
\usepackage{url}
\usepackage{verbatim}
\usepackage{graphicx}
\usepackage{cite}
\usepackage{listings}
\usepackage{amsmath,amsfonts,color}
\usepackage{graphicx}
\usepackage{amsfonts}
\usepackage{array}
\usepackage{amssymb}
\usepackage{helvet}
\usepackage{color}
\usepackage{bm}
\usepackage{float}
\usepackage{cite}
\usepackage{multirow}
\usepackage{slashbox}
\usepackage{amssymb}
\usepackage[caption=false,font=normalsize,labelfont=sf,textfont=sf]{subfig}
\usepackage{amsmath,amsfonts}
\usepackage{algorithmic}
\usepackage{algorithm}
\usepackage{textcomp}
\usepackage{stfloats}
\usepackage{url}
\usepackage{verbatim}
\usepackage{graphicx}
\usepackage{cite}
\usepackage{amssymb}
\usepackage{makecell}
\usepackage{mathrsfs}

\newcommand{\RNum}[1]{\uppercase\expandafter{\romannumeral #1\relax}}

\newcounter{MYtempeqncnt}

\begin{document}

\title{6D Motion  Parameters Estimation   in Monostatic Integrated Sensing and Communications   System}
\author{Hongliang Luo, Feifei Gao, Fan Liu, and Shi Jin
\thanks{H. Luo and F. Gao are with Department of Automation, Tsinghua University, Beijing 100084, China (email: luohl23@mails.tsinghua.edu.cn; feifeigao@ieee.org).}
\thanks{F. Liu is with the Department of Electrical and Electronic Engineering,
Southern University of Science and Technology, Shenzhen 518055, China
(e-mail: liuf6@sustech.edu.cn).}
\thanks{S. Jin is with the National Mobile Communications Research Laboratory, Southeast University, Nanjing 210096, China (e-mail:
jinshi@seu.edu.cn).
}
}



\maketitle

\begin{abstract}
In this paper, we propose a novel scheme to estimate  the six dimensional (6D) motion parameters of dynamic target  for monostatic integrated sensing and communications (ISAC) system.
We first provide a generic ISAC framework for dynamic target sensing based on massive multiple input and multiple output (MIMO) array. 
Next, we derive the relationship between the sensing  channel of ISAC base station (BS) and the 6D motion parameters of dynamic target.
Then,  we employ the  array signal processing methods to estimate the horizontal angle, pitch angle, distance, and \emph{virtual velocity} of   dynamic target. 
Since the virtual velocities observed by different antennas are different, we adopt plane  fitting to estimate the  dynamic target's radial velocity, horizontal angular velocity, and pitch angular velocity from these virtual velocities. 
Simulation results  demonstrate the effectiveness of the proposed 6D motion parameters estimation scheme,   which also confirms a new finding  that \emph{one single BS with massive MIMO array is capable of estimating the horizontal angular velocity and pitch angular velocity  of dynamic target.} 
\end{abstract}

\begin{IEEEkeywords}
6D  sensing, angular velocity estimation, integrated sensing and communications, dynamic target sensing.
\end{IEEEkeywords}

\section{Introduction}

In the past decade, the integration of wireless communications and radar sensing has promoted the researches on
 dual functions radar communications (DFRC) systems\cite{8828023,8999605,9540344}. 
With the  expansion of the sensing category, integrated sensing and communications (ISAC) that incorporates more diverse sensing  technologies based on DFRC has been recognized as a promising air-interface technology for the next-generation wireless networks\cite{202310141,9040264,9606831}.
Since ISAC allows sensing systems and communications systems to share spectrum resources, as well as create more business scenarios for wireless systems,
 it has  been officially approved  by International Telecommunication Union (ITU) as one of the key usage scenarios for the sixth generation (6G) mobile communications\cite{itu,a11221,9755276}.

The ultimate functionality  of sensing is to build the mapping relationship from  real physical world to  digital twin world,
where the former can be decomposed into  
static environment (such as roads, buildings, and trees) and dynamic targets (such as pedestrians, vehicles, and unmanned aerial vehicles). 
Therefore, realizing static environment reconstruction  and dynamic target sensing   becomes   one consensus in the research field of ISAC\cite{2024arXiv240519925L}.
Specifically, 
dynamic target sensing,  as a research focus,   refers to the discovery, detection, parameters estimation, tracking, imaging, and recognition of the target.

Depending on the number and location of base stations (BSs),   ISAC systems can be divided into: 
1) monostatic ISAC system (only one BS in the system); 
2) bistatic ISAC system (two BSs in the system); 
and 
3) multistatic ISAC system (multiple BSs in the system)\footnote{Similarly, static radar systems can also be divided into monostatic radar system, bistatic radar system, and  multistatic radar system.}\cite{9906898,2023arXiv230512994H}.
Among different ISAC architectures, monostatic ISAC system has received tremendous research attention due to its convenient  implementation, as it does not require high-precision synchronization among BSs. Estimating the motion parameters of targets with monostatic ISAC system is a basic issue of dynamic target sensing.
The motion parameters of  target in three-dimensional (3D) space mainly include the 3D position and 3D velocity, which are usually converted into the
\emph{6D motion parameters} relative to  radar  or BS, i.e., 
horizontal angle (HA), pitch angle (PA), distance (Dis), radial velocity (RV), horizontal angular velocity (HAV), and pitch angular velocity (PAV). 
Among them, estimating HA, PA and Dis constitutes  {position estimation}, while estimating RV, HAV and PAV constitutes  {velocity estimation}.

In the early research of radar system, since monostatic radar  has only one transmitting and receiving antenna,    the radial relative motion between the wave source (target) and the single observer (radar) would cause Doppler effect. Thus, monostatic radar with one antenna can only estimate target's Dis and  RV,  which is termed as ``2D radar''. 
With the development of  multiple input and  multiple output (MIMO) technology, monostatic radar  equipped with uniform planar array (UPA) can futher estimate target's HA and PA on the basis of 2D radar. 
Hence the existing  ``4D MIMO radar''  can estimate target's  HA, PA, Dis, and RV\cite{9429942,9913510,2023arXiv230604242H,10477463}.

The counterpart of  monostatic radar, i.e.,   the monostatic ISAC system mainly adopts massive MIMO array and orthogonal frequency division multiplexing (OFDM) modulation to estimate the motion parameters of dynamic targets.  
For example, M.~F.~Keskin~\emph{et.~al.}  utilized multiple signal classification (MUSIC) algorithm to estimate the HA of dynamic target, and then adopted  least-squares (LS) approach to estimate the Dis and RV of the  target in monostatic ISAC system\cite{9529026}.
 X.~Chen~\emph{et.~al.}  proposed a MUSIC based monostatic ISAC system that can  attain high  accuracy in target's HA, PA, Dis, and RV estimation\cite{10048770}.
M.~L.~Rahman~\emph{et.~al.}  proposed a compressive sensing (CS) based ISAC system  to estimate the dynamic target's HA, Dis, and RV\cite{8827589}.
W.~Jiang~\emph{et.~al.}  proposed a model-driven   ISAC scheme, which simultaneously accomplished tasks of  demodulating  uplink communications signals and estimating the Dis/RV of the dynamic target\cite{2023arXiv230715074J}.
R.~Zhang~\emph{et.~al.}  proposed a joint communications  channel  and target parameter estimation scheme based on unified tensor calculation, in which the dynamic target's HA, Dis, and RV are estimated through tensor decomposition\cite{10403776}. 
Z.~Du~\emph{et.~al.} leveraged matched filtering and maximum likelihood estimation to obtain the HA, PA, Dis, and RV of   target, and then used the extended Kalman filtering to track the target\cite{9947033}.
However, it can be noticed that all of these monostatic ISAC works follows the same principle of 4D MIMO radar, and thus  ignore the estimation of target's HAV and PAV.

There do exist some separate studies estimating the HAV of target based on monostatic radar. 
J.~A.~Nanzer~\emph{et.~al.}  first proposed a theoretical method to  estimate the angular velocity of moving object based on spatial interferometry under  monostatic radar  system\cite{5634150,6697333,6711074,6236072}.
X.~Wang~\emph{et.~al.}  extended this work to estimate the HAV of multiple targets \cite{8835771,8936370,9114597}. 
However, these studies only considered the spatial interference effect between two or three antennas, and cannot be directly extended to radar or BS with massive number of antennas. 

In this paper, we propose 
 a novel 6D motion parameters estimation scheme for monostatic ISAC system that can simultaneously estimate the HA, PA, Dis, RV, HAV, and PAV of the dynamic target.
The contributions of this paper are summarized as follows.

\begin{itemize}
	
\item  We propose a generic ISAC framework for dynamic target sensing based on massive MIMO array,  in which the BS adopts multi-beam technology to realize the sensing scanning of  coverage area and the real-time tracking of  the discovered targets while maintaining communications services.

\item We re-examine  the relationship between  6D motion parameters of dynamic target  and  sensing echo channel, based on which we  derive the 6D sensing channel model for  monostatic MIMO-OFDM-ISAC system.

\item We propose a novel 6D motion parameters estimation scheme for dynamic target sensing in monostatic ISAC system.
Specifically, we  employ  the array signal processing technique to estimate dynamic target's  HA, PA, Dis, and virtual velocity. Then we show  that the virtual velocities observed at different antennas are distinct from each other, based on which we utilize plane  fitting method to estimate dynamic target's RV, HAV, and PAV.

\item Simulation results demonstrate the effectiveness of the proposed 6D motion parameters estimation scheme.

\end{itemize}

\begin{figure}[!t]
	\centering
	\includegraphics[width=88mm]{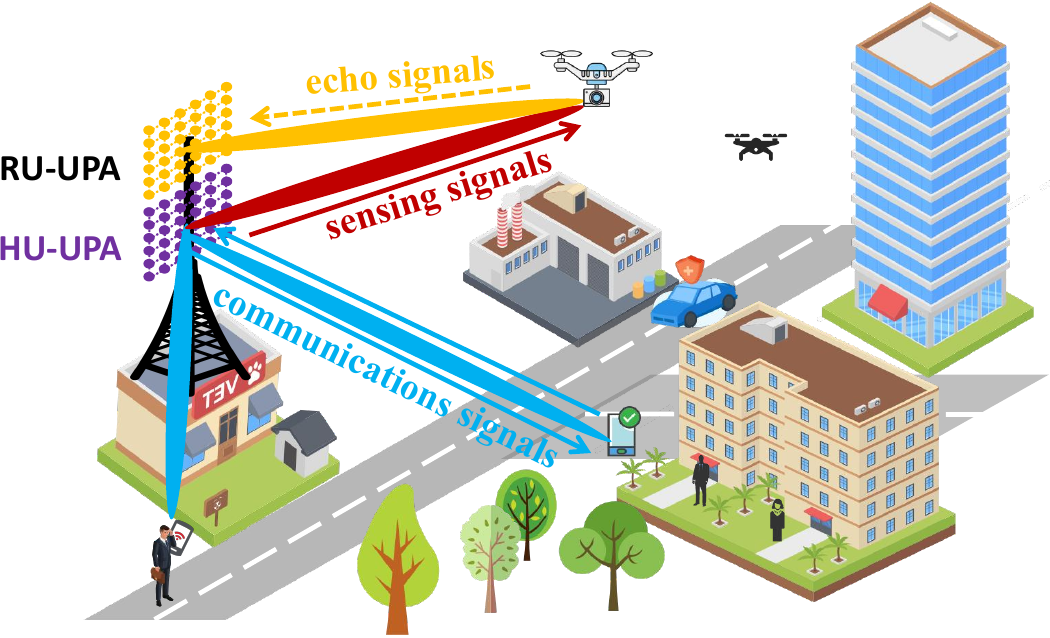}
	\caption{ISAC scene.}
	\label{fig_1}
\end{figure}

The remainder of this paper is organized as follows.
In Section \RNum{2}, we
propose an ISAC framework for dynamic target sensing. 
In Section~\RNum{3}, we derive  the  6D sensing channel of dynamic targets.
In Section~\RNum{4}, we propose 
a novel 6D motion parameters estimation scheme for dynamic target sensing. 
Simulation results and conclusions are given in Section~\RNum{5} and Section~\RNum{6},
respectively.

\emph{Notation}:
Lower-case and upper-case boldface letters $\mathbf{a}$ and $\mathbf{A}$ denote a vector and a matrix;
$\mathbf{a}^T$ and $\mathbf{a}^H$ denote the transpose and the conjugate transpose of vector $\mathbf{a}$, respectively;
$[\mathbf{a}]_n$  denotes the $n$-th element of the vector $\mathbf{a}$;
$[\mathbf{A}]_{i,j}$ denotes the $(i,j)$-th element of the matrix $\mathbf{A}$; $\mathbf{A}[i_1:i_2,:]$ is the submatrix composed of all columns elements in rows $i_1$ to $i_2$ of matrix $\mathbf{A}$;
$\mathbf{A}[:,j_1:j_2]$ is the submatrix composed of all rows elements in columns $j_1$ to $j_2$ of matrix $\mathbf{A}$;
${\rm eig}(\cdot)$ represents the matrix eigenvalue decomposition function;
$\delta (\cdot)$ denotes the Dirac delta function; 
$\otimes$ denotes the Kronecker product;
$\mathbb{R}$ and  $\mathbb{C}$   represent the set of real numbers and the set of complex numbers, respectively;
$\mathcal{R}(a)$ and  $\mathcal{I}(a)$  represent taking the real part and imaginary part of $a$, respectively.

\section{Proposed ISAC Framework for \\ Dynamic Target Sensing}

In this section, we provide the basic architecture of massive MIMO based monostatic ISAC system, and design the generic ISAC   framework for dynamic target sensing.

\subsection{ISAC Scene}

Fig.~1 depicts a massive MIMO based monostatic ISAC system
operating in  mmWave frequency bands with OFDM modulation, which employs only one dual-functional BS for wireless communications and target sensing at the same time.
The BS consists of one \emph{hybrid unit (HU)} and one \emph{radar  unit (RU)}.
HU is responsible for transmitting downlink communications signals and receiving uplink communications signals, as well as transmitting downlink sensing  signals. 
RU is   responsible for receiving   echo signals.

HU and  RU are each equipped with one uniform planar array (UPA) of $N_H=N^{x}_{H}\times N_{H}^z$ and $N_R=N_{R}^x\times N_{R}^z$ antenna elements,  named as
HU-UPA and RU-UPA, respectively. 
Assume that both  HU-UPA and   RU-UPA are   vertically mounted on the 2D plane $y = 0$ at the BS side, as shown in Fig.~2,
while the antenna spacing between the antennas  along  x-axis and z-axis are 
$d_x = d\le \frac{\lambda}{2}$ and 
$d_z = d\le \frac{\lambda}{2}$, respectively,
with $\lambda$ being the wavelength. 
Without loss of generality, we denote the
position of the $n_H$-th antenna element  in  the HU-UPA as $\mathbf{p}_{n_H}=\mathbf{p}_{0_H}+[d\cdot n_{H}^x,0,d\cdot n_{H}^z]^T$, where
$\mathbf{p}_{0_H}$ is the position of the reference  element, 
 $n_{H}^x \in \{0,1,...,N_{H}^x-1\}$ and $n_{H}^z \in \{0,1,...,N_{H}^z-1\}$ are the antenna indices. 
Here we  use two types of index numbers to represent the same antenna,
that is, the $n_H$-th antenna may also be named as the $(n^x_H,n^z_H)$-th antenna.
Similarly, we denote the
position of the $n_R$-th antenna element  in  RU-UPA as $\mathbf{p}_{n_R}= \mathbf{p}_{0_R} + [d\cdot n_R^x,0,d\cdot n_R^z]^T$ with
$n_{R}^x \in \{0,1,...,N^x_{R}-1\}$ and $n_{R}^z  \in \{0,1,...,N^z_{R}-1\}$. The $n_R$-th antenna may also be named as the $(n^x_R,n^z_R)$-th antenna.

\begin{figure}[!t]
	\centering
	\includegraphics[width=75mm]{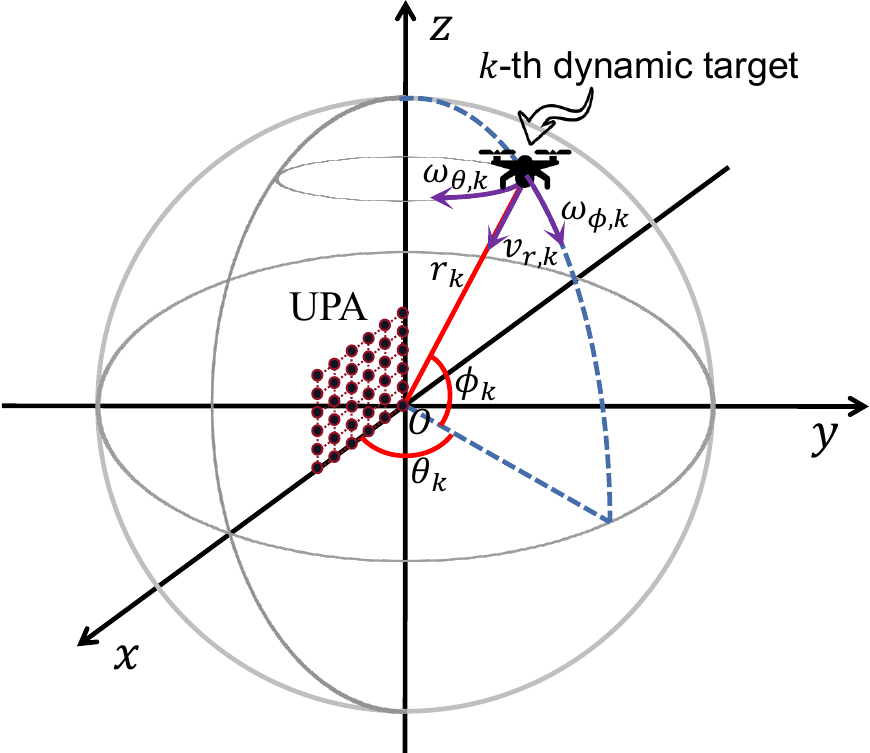}
	\caption{BS's UPA and dynamic target's 6D motion parameters.}
	\label{fig_1}
\end{figure}

We further  assume that HU-UPA and  RU-UPA
are co-located and are parallel to each other, i.e., $\mathbf{p}_{0_H} = \mathbf{p}_{0_R} = [0,0,0]^T$, such that they can see the targets at the same propagation directions\footnote{Since the normal communications and sensing distance is much longer than the protection distance between  HU-UPA and RU-UPA, it can be considered that HU-UPA and RU-UPA are located in the same position.}\cite{9898900}.
We also assume that  HU-UPA employs the  hardware architecture based on  phase shifter (PS) structure, in which a total of $N_{H,RF}\ll N_H$ radio frequency (RF) chains are deployed, and each antenna is connected to one PS  to realize beamforming.  
On the other side, we  assume that  RU-UPA employs the fully-digital receiving array, with each antenna connecting to one RF chain,  to realize super-resolution sensing \cite{10048770,9898900}.

Suppose that the ISAC system emits   OFDM signals with $M$ subcarriers,
where the  lowest  frequency and the subcarrier  interval of OFDM signals are $f_0$ and $\Delta f$, respectively.
Then the transmission bandwidth is $W=M\Delta f$, and 
the frequency of the $m$-th subcarrier  is $f_m=f_0+m\Delta f$, where $m=0,1,...,M-1$. 
Moreover, we consider that an OFDM frame contains $N$ consecutive OFDM symbols, where  
the time interval between the adjacent OFDM symbols is  $T_s = T'_s+T_g$, 
with $T'_s = \frac{1}{\Delta f}$ and $T_g$
being the OFDM symbol duration and guard interval, respectively\cite{4570206}.
Then the starting time of the $n$-th OFDM symbol in one frame is
$t_n=nT_s$, where $n=0,1,...,N-1$. 

\begin{figure*}[!t]
	\centering
	\includegraphics[width=170mm]{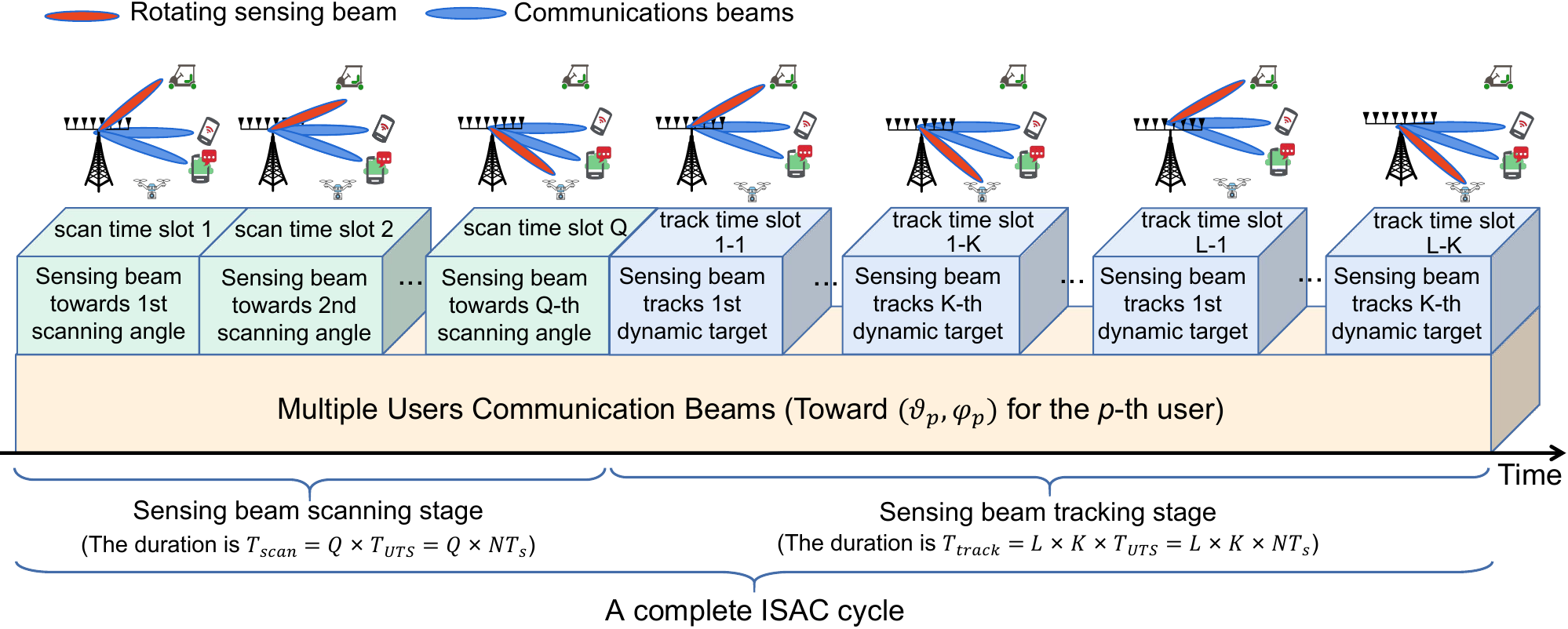}
	\caption{The proposed ISAC framework.}
	\label{fig_1}
\end{figure*}

Let us employ the spherical coordinate $(r,\theta,\phi)$  in 3D space. As shown in Fig.~2, $r$  represents the polar Dis with   range of $r \geq 0$,
$\theta$ represents  the HA with  range of $0^\circ \leq \theta \leq 180^\circ$, and $\phi$  represents the PA with   range of $-90^\circ \leq \phi \leq 90^\circ$. 
Moreover, the spherical coordinate $(r,\theta,\phi)$ 
may be translated to its Cartesian counterpart  $(x,y,z)$ through 
\begin{align}
	x=r\cos \phi \cos \theta,
	\quad y=r\cos \phi \sin \theta,
	\quad z=r\sin \phi \label{3}.
\end{align}

Since the BS is located at the origin of the coordinate system, we  denote the service area of BS as 
$\{(r,\theta,\phi)|r_{min}\leq r \leq r_{max},\theta_{min}\leq \theta \leq \theta_{max},\phi_{min}\leq \phi \leq \phi_{max}\}$. 

Suppose that there are $P$ communications users, $K$  dynamic targets, as well as the widely distributed static environment within this service area.
Denote the spherical coordinate position of the $p$-th user as $(R_p,\vartheta_p,\varphi_p)$,
which can be assumed known and stationary to  BS since they can  be easily obtained through  user reporting  or other techniques\cite{9782674,10271123,9013639}.
Besides, as shown in Fig.~2,  we assume that the 6D motion parameters of the $k$-th dynamic target are
$\{r_k, \theta_k, \phi_k, v_{r,k}, \omega_{\theta,k}, \omega_{\phi,k}\}$,
in which $r_k$, $\theta_k$, $\phi_k$, $v_{r,k}$,  
$\omega_{\theta,k}$ and $\omega_{\phi,k}$
 represent the Dis, HA, PA, RV, HAV, and PAV of the $k$-th dynamic target, respectively. 
Without loss of  generality,  during the  motion process, the direction in which the Dis value decreases, the direction in which the HA value decreases, and the direction in which the PA value decreases  are taken as the positive directions for RV, HAV, and PAV, respectively.

\subsection{The Proposed ISAC  Framework}

The task of  ISAC system is to sense  all $K$ dynamic targets from  the static environment while serving $P$  communications users. As depicted  in Fig.~3,
the proposed ISAC framework consists of two stages: \emph{sensing beam scanning (SBS) stage} and  \emph{sensing beam tracking (SBT) stage}.

For the aspect of communications,   BS continuously generates $P$ communications beams  towards $P$    users
to  maintain  communications service during both SBS stage and SBT stage. 
For the aspect of sensing,  BS 
generates one  sensing beam that can  scan 
the  service area during  SBS stage, through which the  BS may detect the  targets
and estimate their parameters\cite{10477890}.
Next,  BS generates one sensing beam  to track all $K$  dynamic targets in a time division manner during  SBT stage,
that is, the sensing beam may sequentially illuminate each of the $K$ targets and continuously track them.

Without loss of generality, assume that $K$ dynamic targets stay at  different physical  directions. For each direction,   BS  adopts one OFDM frame, i.e., $N$ consecutive OFDM symbols,  to realize dynamic target motion parameters estimation. 
Thus we divide the working cycle of  ISAC system into a large number of unit time slots (UTSs), and each UTS occupies one OFDM frame, that is, each UTS lasts for a time duration of $T_{UTS} =  N\cdot T_s$.
As shown in Fig.~3, we divide  SBS stage into $Q$ scanning time slots (STSs), and  each STS occupies one UTS.
Then  BS needs to scan the $q$-th sensing scanning  angle  direction in the $q$-th STS.
Besides,   we  divide  SBT stage  into $L\times K$ tracking time slots (TTSs),  and each TTS occupies one UTS. 
Then  BS tracks the $k$-th target  for  the $l$-th time within the $(l,k)$-th TTS.

For both STS and TTS, BS needs to simultaneously generate $P$ communications beams pointing to the users and one sensing beam pointing to the current sensing direction (CSD)   $(\xi,\eta)$  from   HU-UPA.
We assume that the transmission power of  BS is $P_t$,
the energy of sensing beam is  $\rho_{\xi,\eta}^s P_t$,
and the energy of the $p$-th communications beam is $\rho_{\xi,\eta}^{c,p} P_t$, where $p=1,2,...,P$, $\rho_{\xi,\eta}^s\in [0,1]$ and $\rho_{\xi,\eta}^{c,p}  \in [0,1]$ are the power distribution coefficients used for the CSD $(\xi,\eta)$ with $\rho_{\xi,\eta}^s+\sum_{p=1}^{P}\rho_{\xi,\eta}^{c,p}=1$. 
Then the transmission signals from  HU-UPA  on the $m$-th subcarrier of the $n$-th  symbol in the UTS for  CSD $(\xi,\eta)$ can be represented as
\begin{equation}
	\begin{split}
		\begin{aligned}
			\label{deqn_ex1a}
& \mathbf{x}_{\xi,\eta,n,m} = \sum_{p=1}^{P}\mathbf{w}_{c,p,\xi,\eta}s^{c,p,\xi,\eta}_{n,m}+\mathbf{w}_{s,\xi,\eta}s^{s,\xi,\eta}_{n,m}
\\& =  \sum_{p=1}^{P}\sqrt{\frac{\rho_{\xi,\eta}^{c,p}P_{t}}{N_H}}\mathbf{a}_{H}\!\left(\Psi(\vartheta_p,\varphi_p), \Omega(\vartheta_p,\varphi_p)\right)s^{c,p,\xi,\eta}_{n,m}\\&\quad + \sqrt{\frac{\rho_{\xi,\eta}^sP_{t}}{N_H}}\mathbf{a}_{H}\!\left(\Psi({\xi,\eta}),\Omega(\xi,\eta)\right)s^{s,\xi,\eta}_{n,m},
		\end{aligned}
	\end{split}
\end{equation}
where
\begin{align}
\Psi(\theta,\phi) &= \cos \phi \cos \theta \label{1},\\
\Omega(\theta,\phi)&=\sin\phi \label{2}.
\end{align}
Define $(\Psi(\theta,\phi),\Omega(\theta,\phi))$ as the {spatial-domain direction (SDD)} corresponding to  physical  direction  $(\theta,\phi)$. 
We will refer to $\Psi(\theta,\phi)$ and $\Omega(\theta,\phi)$ as the horizontal spatial-domain direction and the pitch spatial-domain direction, respectively.
Besides, 
 $\mathbf{w}_{c,p,\xi,\eta}=\sqrt{\frac{\rho_{\xi,\eta}^{c,p}P_t}{N_H}}\mathbf{a}_{H}\!\left(\Psi(\vartheta_p,\varphi_p), \Omega(\vartheta_p,\varphi_p)\right)$ and $\mathbf{w}_{s,\xi,\eta}=\sqrt{\frac{\rho_{\xi,\eta}^sP_t}{N_H}}\mathbf{a}_{H}\!\left(\Psi({\xi,\eta}),\Omega(\xi,\eta)\right)$ are  communications beamforming vector for the $p$-th user  and  sensing beamforming vector for   CSD $(\xi,\eta)$, respectively, while  $\mathbf{a}_{H}(\Psi,\Omega)$ is the array   steering vector  of HU-UPA with the form\cite{7523373}
\begin{equation}
	\begin{split}
		\begin{aligned}
			\label{deqn_ex1a}
\mathbf{a}_{H}(\Psi,\Omega)\! =
\mathbf{a}_{H}^x(\Psi)\otimes \mathbf{a}_{H}^z(\Omega)
	 \in\! \mathbb{C}^{N_H\times 1},
		\end{aligned}
	\end{split}
\end{equation}
where $\otimes$ denotes the Kronecker product, and 
\begin{align}
\!\!\mathbf{a}_{H}^x(\Psi)&\!=\![1,e^{j\frac{2\pi f_0d\Psi}{c}},...,e^{j\frac{2\pi f_0d\Psi}{c}(N_{H}^x-1)}]^T \!\!\in\! \mathbb{C}^{N_{H}^x\times 1} \label{1},\\
\!\!\mathbf{a}_{H}^z(\Omega)&\!=\! [1,e^{j\frac{2\pi f_0d\Omega}{c}},...,e^{j\frac{2\pi f_0d\Omega}{c}(N_{H}^z-1)}]^T \!\! \in\! \mathbb{C}^{N_{H}^z\times 1} \label{2}.
\end{align}
Moreover,  $s^{c,p,\xi,\eta}_{n,m}$ and $s^{s,\xi,\eta}_{n,m}$ are  communications signals for the $p$-th user and  
sensing detection signals, respectively.

With (2),  BS may realize both communications function and sensing function by optimizing $\{\rho_{\xi,\eta}^s,\rho_{\xi,\eta}^{c,1} ,...,\rho_{\xi,\eta}^{c,P}\}$ during each  UTS, 
which may be conceived through the power allocation strategy  in 
 \cite{10477890}.
To that end,   we will only focus on   dynamic target sensing    problem while omit the design of communications function in this work, as did in \cite{2023arXiv231213154H,2024arXiv240509443Z}. 
Consequently, there is always one beam  towards the CSD $(\xi,\eta)$, and  we can rewrite the transmission signals  on the $m$-th subcarrier of the $n$-th OFDM symbol for the CSD $(\xi,\eta)$ as
\begin{equation}
	\begin{split}
		\begin{aligned}
			\label{deqn_ex1a}
\grave{\mathbf{x}}_{\xi,\eta,n,m} =  \sqrt{\frac{\grave{\rho}_{\xi,\eta}P_t}{N_H}}\mathbf{a}_{H}\left(\Psi({\xi,\eta}),\Omega(\xi,\eta)\right)s^{\xi,\eta}_{n,m},
		\end{aligned}
	\end{split}
\end{equation}
where $\grave{\rho}_{\xi,\eta}P_t$ is  the power allocated to  $(\xi,\eta)$ direction, and $s^{\xi,\eta}_{n,m}$ is the  signal transmitted through  $(\xi,\eta)$ direction.

\section{6D Sensing Channel  of  Monostatic ISAC System}

In this section, we first derive  the basic principles of the sensing  channel for monostatic  ISAC system, and then propose  the modified 6D sensing channel of the dynamic target.

\subsection{The Basic Principles of Sensing Channel}

In one UTS, the BS transmits   the   detection signals through HU-UPA at the beginning of sensing, which will be reflected by dynamic targets. Then  RU-UPA   will receive the sensing echo signals. 
Let us define  the path from the $n_H$-th antenna of  HU-UPA to the $k$-th dynamic target and then back to the $n_R$-th antenna of RU-UPA as the $(n_H,k,n_R)$-th propagation path. 
Define
\begin{equation}
	\begin{split}
		\begin{aligned}
			\label{deqn_ex1a}
\tau^{n}_{n_H\!,k,n_R}\!=\frac{D^{n}_{k,n_H}+D^{n}_{k,n_R}}{c}
		\end{aligned}
	\end{split}
\end{equation} 
as the time delay of the $(n_H,k,n_R)$-th propagation path at the $n$-th OFDM symbol  during one UTS,
where $c$ represents the speed of light, $D^{n}_{k,n_H}$ is the distance between the $n_H$-th antenna of HU-UPA and the $k$-th  target at the $n$-th OFDM symbol, and $D^{n}_{k,n_R}$ is the distance between the $n_R$-th antenna of RU-UPA and the $k$-th  target at the $n$-th OFDM symbol.

\begin{figure*}[!b]
	\normalsize
	\setcounter{MYtempeqncnt}{\value{equation}}
	\vspace*{4pt}
	\hrulefill 
	\setcounter{equation}{40}
	\begin{equation}
		\begin{split}
			\begin{aligned}
				\label{deqn_ex1a}
D_{k,n_H}^n &= \sqrt{(x_{k,n}-n_{H}^xd)^2+(y_{k,n}-0)^2+(z_{k,n}-n_{H}^zd)^2} \\&= \sqrt{(r_{k,n}\cos \phi_{k,n} \cos \theta_{k,n}-n_{H}^xd)^2+(r_{k,n}\cos \phi_{k,n} \sin \theta_{k,n})^2+(r_{k,n}\sin \phi_{k,n}-n_{H}^zd)^2}\\&\approx 
r_{k,n} - (n^x_Hd\cos \phi_{k,n}\cos \theta_{k,n}+n^z_Hd\sin \phi_{k,n}).
			\end{aligned}
		\end{split}\tag{14}
	\end{equation}
	\setcounter{equation}{\value{MYtempeqncnt}}
\end{figure*}

Suppose that the signal transmitted by the
$n_H$-th antenna of  HU-UPA is $s(t)$,
 and the corresponding passband signal is  $\mathcal{R}\{s(t)e^{j2\pi f_0t}\}$. 
Then the echo signal will be a delayed version of the transmitting signal with amplitude attenuation.
Specifically, the passband echo signal received by the $n_R$-th antenna of RU-UPA 
through the $(n_H,k,n_R)$-th propagation path
at the $n$-th OFDM symbol  is
\begin{equation}
	\begin{split}
		\begin{aligned}
			\label{deqn_ex1a}
\mathcal{R}\left\{\alpha_{n_H,k,n_R}^n s\left(t-\tau^{n}_{n_H,k,n_R}\right)e^{j2\pi f_0\left(t-\tau^{n}_{n_H,k,n_R}\right)}\right\},
		\end{aligned}
	\end{split}
\end{equation} 
where $\alpha_{n_H,k,n_R}^n\approx \alpha_{k}=\sqrt{\frac{\lambda^2}{(4\pi)^3 (r_{k})^4}}\sigma_{k}$ is the  channel fading factor,  and 
$\sigma_{k}$  is the radar cross section (RCS) of the $k$-th dynamic target.
Without loss of  generality, we  assume that the RCS follows the   Swerling $\rm \uppercase\expandafter{\romannumeral1}$ model\cite{2024arXiv240417462W,1057561}. 
Then the corresponding baseband echo signal is
$\alpha_{k} s(t-\tau^{n}_{n_H,k,n_R})e^{-j2\pi f_0\tau^{n}_{n_H,k,n_R}}$, 
and  the corresponding time-domain baseband  equivalent channel is
\begin{equation}
\begin{split}
\begin{aligned}
\label{deqn_ex1a}
h^{n}_{n_H,k,n_R}(t) = \alpha_k e^{-j2\pi f_0\tau^{n}_{n_H,k,n_R}} \delta \left(t-\tau^{n}_{n_H,k,n_R}\right),
\end{aligned}
\end{split}
\end{equation}
where $\delta (\cdot)$ denotes the Dirac delta function. The  frequency-domain baseband channel response can be obtained via the Fourier transform of (11), i.e.,
\begin{equation}
	\begin{split}
		\begin{aligned}
		\label{deqn_ex1a}
h^{n,F}_{n_H,k,n_R}(f) = \alpha_k e^{-j2\pi (f_0+f)\tau^{n}_{n_H,k,n_R}}.
		\end{aligned}
	\end{split}
\end{equation}
Thus the frequency-domain sensing echo channel  of the $(n_H,k,n_R)$-th propagation path    on the $m$-th subcarrier at the $n$-th OFDM symbol is 
\begin{equation}
	\begin{split}
		\begin{aligned}
			\label{deqn_ex1a}
 h^{n,m}_{n_H\!,k,n_R} \!\!=\! \alpha_k    e^{-j2\pi f_m\tau^{n}_{n_H\!,k,n_R}}  \!=\!\alpha_k e^{\!-j2\pi f_m \frac{D^{n}_{k,n_H}\!+D^{n}_{k,n_R}}{c}}.
		\end{aligned}
	\end{split}
\end{equation}

Since the positions of the antennas are fixed, $D^{n}_{k,n_H}$ and $D^{n}_{k,n_R}$  would change as the position of the $k$-th target changes within one frame of OFDM symbols.
Let us define the 6D motion parameters of the $k$-th dynamic target at the $n$-th OFDM symbol as 
$\{r_{k,n}, \theta_{k,n}, \phi_{k,n}, v_{r,k,n}, \omega_{\theta,k,n}, \omega_{\phi,k,n}\}$, where
$\{r_{k,0}, \theta_{k,0}, \phi_{k,0}, v_{r,k,0}, \omega_{\theta,k,0}, \omega_{\phi,k,0}\}=\{r_{k}, \theta_{k}, \phi_{k}, v_{r,k}, \omega_{\theta,k}, \omega_{\phi,k}\}$. Besides,  
 $(r_{k,n}, \theta_{k,n}, \phi_{k,n})$  is the spherical coordinate position of the $k$-th target at the $n$-th OFDM symbol, and  the corresponding Cartesian coordinate position can be obtained through (1) as 
$(x_{k,n},y_{k,n},z_{k,n})=(r_{k\!,n}\!\cos\! \phi_{k,n} \!\cos\! \theta_{k,n},r_{k,n}\!\cos\! \phi_{k,n}\! \sin\! \theta_{k,n},r_{k,n}\!\sin\! \phi_{k,n})$. Note that the Cartesian coordinate position of the $n_H$-th antenna of HU-UPA is $(d\cdot n_{H}^x,0,d\cdot n_{H}^z)$. Then 
the Taylor expansion approximation of 
$D^{n}_{k,n_H}$ can be expressed as (14) at the bottom of this page. 
Similarly, $D^{n}_{k,n_R}$ can be approximated as $D^{n}_{k,n_R} \approx  r_{k,n} - (n^x_Rd\cos \phi_{k,n}\cos \theta_{k,n}+n^z_Rd\sin \phi_{k,n})$. 
Based on (3) and (4), there are $\cos \phi_{k,n}\!\cos \theta_{k,n}=\Psi(\theta_{k,n},\phi_{k,n})$ and $\sin \phi_{k,n} = \Omega (\theta_{k,n},\phi_{k,n})$.
For clarity, we abbreviate that 
\begin{align}
	\Psi_{k,n} &= \Psi(\theta_{k,n},\phi_{k,n})=\cos \phi_{k,n}\!\cos \theta_{k,n} \tag{15},\\
	\Omega_{k,n}&=\Omega (\theta_{k,n},\phi_{k,n})=\sin \phi_{k,n} \tag{16}.
\end{align}
Then there is
\begin{equation}
	\begin{split}
		\begin{aligned}
			\label{27}
 \tau^{n}_{n_{H},k,n_{R}} \!=\!\frac{ 2r_{k,n}\!\!-\!(n^x_{\!H}\!+\!n^x_{\!R})d\Psi_{k,n}\!-\!(n^z_{H}\!+\!n^z_{R})d\Omega_{k,n}}{c}.
		\end{aligned}
	\end{split}\tag{17}
\end{equation}
Based on (17), Eq. (13) can be rewritten as
\begin{equation}
	\begin{split}
		\begin{aligned}
			\label{27}
\!\! h^{n,m}_{n_H\!,k,n_R} \!=\! \alpha_k e^{-j2\pi f_m \! \frac{2r_{k,n}\!-(n^x_H+n^x_R)d\Psi_{k,n}\!-(n^z_H+n^z_R)d\Omega_{k,n}}{c}}.
		\end{aligned}
	\end{split}\tag{18}
\end{equation}
Eq. (18) describes the underlying form of the sensing echo channel in ISAC system, which includes all the motion parameters of the dynamic target.

\subsection{The Proposed 6D Sensing Channel}

The true velocity of a dynamic target in 3D space is a directional 3D velocity vector, and thus can be transformed into the  representations of RV, HAV, and PAV in the spherical coordinate system with BS being the origin.
During the process of dynamic target sensing,
due to the very short duration of one frame of OFDM symbols, it is generally assumed that the motion velocity of the dynamic target does not change within one frame of OFDM symbols\cite{9529026,10048770,8827589,2023arXiv230715074J,10403776,9947033}. 
Then the  motion parameters of the $k$-th dynamic target at the $n$-th OFDM symbol are  
\begin{align}
	v_{r,k,n}&=v_{r,k,0}=v_{r,k}
	\tag{19},\\
	\omega_{\theta,k,n}&=\omega_{\theta,k,0}=\omega_{\theta,k}
	\tag{20},\\
	\omega_{\phi,k,n}&=\omega_{\phi,k,0}=\omega_{\phi,k} \tag{21},\\
	r_{k,n}&= r_{k,0} - v_{r,k}nT_{s}=r_{k} - v_{r,k}nT_{s}  \tag{22},\\
	\theta_{k,n}&=\theta_{k,0}-\omega_{\theta,k}nT_{s}=\theta_{k}-\omega_{\theta,k}nT_{s}
	\tag{23},\\
\phi_{k,n}&=\phi_{k,0} - \omega_{\phi,k}nT_{s}=\phi_{k} - \omega_{\phi,k}nT_{s}.\tag{24}
\end{align}
Based on  (23) and (24),   Eq. (15) and Eq. (16) can be rewritten as
\begin{align}
\Psi_{k,n}&=
\cos (\phi_{k} - \omega_{\phi,k}nT_{s})\!\cos (\theta_{k}-\omega_{\theta,k}nT_{s}), \tag{25}\\
\Omega_{k,n}&=\sin (\phi_{k} - \omega_{\phi,k}nT_{s}).
	\tag{26}
\end{align}
Based on (22), (25) and (26), 
Eq. (18) can be rewritten as 
 \begin{equation}
 	\begin{split}
 		\begin{aligned}
 			\label{27}
 			&\!\!\!h^{n,m}_{n_H\!,k,n_R} \!=\! \alpha_k e^{-j2\pi f_m \! \frac{2r_{k,n}\!-(n^x_H+n^x_R)d\Psi_{k,n}\!-(n^z_H+n^z_R)d\Omega_{k,n}}{c}}\\&\!\!\!\!\!\!=
 			\!\alpha_k e^{\!-\!j\!4\pi\! f_m\!\frac{r_{k}}{c}} \!e^{j4\pi\! f_m\!\frac{v_{r\!,k}nT_s}{c}}
 			\!e^{j2\pi\! f_m\!\frac{(n^x_H\!+\!n^x_R)d\Psi_{k,n}+(n^z_H\!+\!n^z_R)d\Omega_{k,n}}{c}}.
 		\end{aligned}
 	\end{split}\tag{27}
 \end{equation}
In narrowband OFDM system,  Doppler squint effect and beam squint effect are typically negligible\cite{10433130}.  Thus there is 
\begin{equation}
	\begin{split}
		\begin{aligned}
			&h^{n,m}_{n_H,k,n_R}  = \alpha_k e^{-j 4\pi f_m \frac{r_k}{c}}
			e^{j4\pi f_0 \frac{ v_{r,k}nT_{s}}{c}} \times
			\\ & \kern 60pt 
			e^{j2\pi f_0\frac{(n^x_H+ n^x_R)d\Psi_{k,n} +(n^z_H+ n^z_R)d\Omega_{k,n} }{c}}\\&=
			\alpha_k e^{-j 4\pi f_m \frac{r_k}{c}}
			e^{j4\pi f_0 \frac{ v_{r,k}nT_{s}}{c}} \times
			\\ & \kern 4pt 
e^{j2\pi\! f_0\!\frac{(\!n^x_H\!+\! n^x_R)d\!\cos (\!\phi_{k} \!- \omega_{\phi\!,k}\!n\!T_{s}\!)\!\cos (\theta_{k}\!-\!\omega_{\theta\!,k}\!nT_{s}\!) +(n^z_H\!+ n^z_R\!)d\!\sin (\phi_{k}\! - \omega_{\phi\!,k}\!nT_{s}) }{c}}.
		\end{aligned}
	\end{split}\tag{28}
\end{equation}

Define  $\mathbf{H}_{k,n,m} \in \mathbb{C}^{N_R\times N_H}$
as the  frequency-domain sensing echo channel matrix on the $m$-th subcarrier at the $n$-th OFDM symbol   from HU-UPA to the $k$-th dynamic  target and then back to RU-UPA, whose $(n_R,n_H)$-th element is
$[\mathbf{H}_{k,n,m}]_{n_R,n_H} = h^{n,m}_{n_H,k,n_R}$. 
Based on (28), the matrix $\mathbf{H}_{k,n,m}$ can be decomposed as 
\begin{equation}
	\begin{split}
		\begin{aligned}
			\label{deqn_ex1a}
\mathbf{H}_{k,n,m} \!\!\!=\!\! \alpha_{k}   e^{\!-\!j\!4\pi\! f_m \!  \frac{r_{k}}{c}}\!
			e^{j4\pi f_0\!\frac{v_{r\!,k}\!n\!T_s}{c}} \! 
			\mathbf{a}_{\!R}(\Psi_{\!k,n}\!,\!\Omega_{k,n})
			\mathbf{a}^T_{\!H}(\Psi_{\!k,n}\!,\!\Omega_{k,n}),
		\end{aligned}
	\end{split}\tag{29}
\end{equation}
where $\mathbf{a}_{R}(\Psi,\Omega)$ is the array  steering vector for the  SDD  $(\Psi,\Omega)$ of RU-UPA with the form
\begin{equation}
	\begin{split}
		\begin{aligned}
			\label{deqn_ex1a}
			\mathbf{a}_{R}(\Psi,\Omega)\! =
			\mathbf{a}_{R}^x(\Psi)\otimes \mathbf{a}_{R}^z(\Omega)
			\in\! \mathbb{C}^{N_R\times 1}.
		\end{aligned}
	\end{split}\tag{30}
\end{equation}
Here $\otimes$ denotes the Kronecker product, and 
\begin{align}
	\!\!\mathbf{a}_{R}^x(\Psi)&\!=\![1,e^{j\frac{2\pi f_0d\Psi}{c}},...,e^{j\frac{2\pi f_0d\Psi}{c}(N_{R}^x-1)}]^T \!\!\in\! \mathbb{C}^{N_{R}^x\times 1} \tag{31},\\
	\!\!\mathbf{a}_{R}^z(\Omega)&\!=\! [1,e^{j\frac{2\pi f_0d\Omega}{c}},...,e^{j\frac{2\pi f_0d\Omega}{c}(N_{R}^z-1)}]^T \!\! \in\! \mathbb{C}^{N_{R}^z\times 1} \tag{32}.
\end{align}
Eq. (29) describes the basic form of the 6D sensing channel for the $k$-th dynamic target. 

It should be noticed that the  ISAC system's ability to sense the velocity of dynamic target  is attributed to the Doppler effect, and   the small positional changes    within $N$ consecutive and short OFDM symbol times ($N$ consecutive observations) are essential to the Doppler effect. 
As can be seen from Eq. (29), all the RV, HAV, and PAV of the  dynamic target decomposed from the true velocity vector in 3D space will cause slight positional changes within $N$ consecutive OFDM symbol observations, resulting in Doppler effect.
Hence $\mathbf{H}_{k,n,m}$ includes the HA, PA, Dis, RV, HAV, and PAV of the $k$-th dynamic target, i.e., the 6D motion parameters  of the $k$-th dynamic target.

\textbf{Remark}: The existing researches on  ISAC system follows the principle of 4D MIMO radar and ignore the HAV  and PAV  parameters of the target \cite{9529026,10048770,8827589,2023arXiv230715074J,10403776,9947033}, that is,  the motion parameters of the dynamic target are simplified as  $\{r_k, \theta_k, \phi_k, v_{r,k}, \omega_{\theta,k}=0, \omega_{\phi,k}=0\}$.
Then the 6D sensing channel in Eq. (29) degenerates into the existing 4D sensing channel, which can be represented as
\begin{equation}
	\begin{split}
		\begin{aligned}
			\label{deqn_ex1a}
\mathbf{H}^{4D}_{k,n,m} \!\!=\! \alpha_k  e^{\!-j\!4\pi\! f_m \!  \frac{r_{k}}{c}}\!
e^{j4\pi f_0\!\frac{v_{r\!,k}\!n\!T_s}{c}} \! 
\mathbf{a}_{\!R}\!(\!\Psi_{k}^{4D},\Omega_{k}^{4D}\!)
\mathbf{a}^T_{\!H}(\!\Psi_{k}^{4D},\Omega_{k}^{4D}\!).
		\end{aligned}
	\end{split}\tag{33}
\end{equation}
Here $\Psi_{k}^{4D}\!=\!\cos \phi_{k}\!\cos \theta_{k},   
	\Omega_{k}^{4D}\!=\! \sin \phi_{k}$. 
Since the existing monostatic ISAC studies ignore the angular velocity of the target,     $\mathbf{H}^{4D}_{k,n,m}$   only includes 4 motion parameters of the $k$-th dynamic target, i.e., the HA, PA, Dis, and RV.

Based on (29), the  sensing echo channel of all $K$ dynamic targets on the $m$-th subcarrier of the $n$-th OFDM symbol  can be represented as
\begin{equation}
	\begin{split}
		\begin{aligned}
			\label{deqn_ex1a}
\mathbf{H}^{target}_{n,m} = \sum_{k'=1}^K
\mathbf{H}_{k',n,m}.
		\end{aligned}
	\end{split}\tag{34}
\end{equation}

Since the real physical world is composed of dynamic targets and static environment,  
RU-UPA  will receive both the effective echoes caused by the interested dynamic targets (dynamic target echoes) and the undesired  echoes caused by uninterested background environment (clutter). 
Following \cite{10477890}, we could model the static environmental clutter channel 
on the $m$-th subcarrier at the $n$-th OFDM symbol as
\begin{equation}
	\begin{split}
		\begin{aligned}
			\label{deqn_ex1a}
\!\!\!\!\! \mathbf{H}^{background}_{n,m} \!=\! \!
\sum_{i'=1}^{I'}\!\beta^{\mathfrak{c}}_{i'}
e^{\!-\!j\!4\pi\! f_{\!m} \!  \frac{r^{\mathfrak{c}}_{i'}}{c}}\!
\mathbf{a}_{\!R}(\!\Psi^{\mathfrak{c}}_{i'}\!,\!\Omega^{\mathfrak{c}}_{i'})
\mathbf{a}^T_{\!H}\!(\!\Psi^{\mathfrak{c}}_{i'}\!,\!\Omega^{\mathfrak{c}}_{i'}),
		\end{aligned}
	\end{split}\tag{35}
\end{equation}
where $I'$ is the total number of  clutter scattering units,  $\Psi^{\mathfrak{c}}_{i'}=\cos\phi^{\mathfrak{c}}_{i'}\cos\theta^{\mathfrak{c}}_{i'}$, $\Omega^{\mathfrak{c}}_{i'}=\sin\phi^{\mathfrak{c}}_{i'}$, 
$(r^{\mathfrak{c}}_{i'},\theta^{\mathfrak{c}}_{i'},\phi^{\mathfrak{c}}_{i'})$ is the position of the $i'$-th  clutter scattering unit, $\beta^{\mathfrak{c}}_{i'} = \sqrt{\frac{\lambda^2}{(4\pi)^3 (r^{\mathfrak{c}}_{i'})^4}}\sigma^{\mathfrak{c}}_{i'}$
is the channel fading factor, 
and $\sigma^{\mathfrak{c}}_{i'}$ is the RCS of the $i'$-th   clutter scattering unit that also 
follows the Swerling $\rm \uppercase\expandafter{\romannumeral1}$ model\cite{2024arXiv240417462W}.
Due to the random distribution of clutter scattering units in various directions and distances, when the number of clutter scattering units $I'$ is large enough, $\mathbf{H}^{background}_{n,m}$ can be  considered as a random channel. Therefore, a low-complexity clutter channel generation method is to directly approximate $\mathbf{H}^{background}_{n,m}$ as
\begin{equation}
	\begin{split}
		\begin{aligned}
			\label{deqn_ex1a}
\mathbf{H}^{background}_{n,m} \approx 
\beta^{\mathfrak{c}} \mathbf{H}^{\mathfrak{c}}_{m}, 
		\end{aligned}
	\end{split}\tag{36}
\end{equation}
where $\mathbf{H}^{\mathfrak{c}}_{m}$  is a complex Gaussian matrix with the dimension $N_R \times N_H$, and $\beta^{\mathfrak{c}}$  is the clutter power regulation factor.
Eq. (35) and Eq. (36) indicate that since the static  clutter scattering unit remains stationary for $N$ OFDM symbol times, $\mathbf{H}^{background}_{n,m}$ would remain unchanged for $N$  symbols.

Based on (34) and (36), the overall sensing echo channel of  dynamic targets and static environment on the $m$-th subcarrier at the $n$-th OFDM symbol  can be represented as 
\begin{equation}
	\begin{split}
		\begin{aligned}
			\label{deqn_ex1a}
\mathbf{H}^{sensing}_{n,m} = 			\mathbf{H}^{target}_{n,m} +\mathbf{H}^{background}_{n,m}.
		\end{aligned}
	\end{split}\tag{37}
\end{equation}

\begin{figure*}[!t]
	\normalsize
	\setcounter{MYtempeqncnt}{\value{equation}}
	\vspace*{4pt}
	\setcounter{equation}{40}
	\begin{equation}
		\begin{split}
			\begin{aligned}
				\label{deqn_ex1a}
\hat{\mathbf{y}}_{k\!,n\!,m} \!\!=\!\!\!\!  \sum_{k'=1}^K \! \! 
\left[ \! \alpha_{k'}  e^{\!-\!j4\pi\! f_m \!  \frac{r_{k'}}{c}}\!
e^{j4\pi \!f_0\!\frac{v_{r,k'}nT_s}{c}}\!
\mathbf{a}_{\!R}(\!\Psi_{k',n},\Omega_{k',n}\!)
\mathbf{a}^T_{\!H}(\!\Psi_{k',n},\Omega_{k',n}\!)\!  \sqrt{\!\frac{\grave{\rho}_{\theta_{k}, \phi_{k}}\!P_t}{N_H}}\!
\mathbf{a}^*_{\!H}\!(\!\cos \! {\phi}_{k} \cos {\theta}_{k},\sin\! {\phi}_{k})\!\right]
+\hat{\mathbf{n}}_{k,n\!,m}.
			\end{aligned}
		\end{split}\tag{42}
	\end{equation}
	\begin{equation}
		\begin{split}
			\begin{aligned}
				\label{deqn_ex1a}
\hat{\mathbf{y}}_{k\!,n,m} &\!\!=\!\!\!\!  \sum_{k'=1}^K \! \! 
\left[ \! \alpha_{k'} e^{\!-\!j4\pi\! f_m \!  \frac{r_{k'}}{c}}\!
e^{j4\pi \!f_0\!\frac{v_{r,k'}n\!T_s}{c}}\!
\mathbf{a}_{\!R}(\!\Psi_{k',n},\Omega_{k',n}\!)
\mathbf{a}^T_{\!H}(\!\Psi_{k',n},\Omega_{k',n}\!)\!  \sqrt{\!\frac{\grave{\rho}_{\theta_{k}, \phi_{k}}\!P_t}{N_H}}\!
\mathbf{a}^*_{\!H}\!(\!\Psi_{k},\Omega_{k}\!)\!\right]
\!+\!  \hat{\mathbf{n}}_{k,n,m}
\\& \!\! = \alpha_{k}  e^{\!-\!j4\pi\! f_m \!  \frac{r_{k}}{c}}\!
e^{j4\pi \!f_0\!\frac{v_{r,k}n\!T_s}{c}}\!
\mathbf{a}_{\!R}(\!\Psi_{k,n},\Omega_{k,n}\!)
\mathbf{a}^T_{\!H}(\!\Psi_{k,n},\Omega_{k,n}\!)\!  \sqrt{\!\frac{\grave{\rho}_{\theta_{k}, \phi_{k}}\!P_t}{N_H}}\!
\mathbf{a}^*_{\!H}\!(\!\Psi_{k},\Omega_{k}\!) +\!  \hat{\mathbf{n}}_{k,n,m}
\\&\!\! = \mathscr{G}_{k}
e^{\!-\!j4\pi\! f_m \!  \frac{r_{k}}{c}}\!
e^{j4\pi \!f_0\!\frac{v_{r,k}n\!T_s}{c}}
e^{j\!\frac{\pi\! f_0d(\Omega_{k\!,n}\!\!-\Omega_{k})(N_H^z\!-\!1)}{c}}
e^{j\!\frac{\pi\! f_0d(\Psi_{k,n}\!\!-\Psi_{k})(N_H^x\!-\!1)}{c}}
\mathbf{a}_{\!R}(\!\Psi_{k\!,n},\Omega_{k\!,n}\!) +\!  \hat{\mathbf{n}}_{k,n,m}.
			\end{aligned}
		\end{split}\tag{45}
	\end{equation}
	\begin{equation}
		\begin{split}
			\begin{aligned}
				\label{deqn_ex1a}
& \hat{y}_{k,n^x_R,n^z_R,n,m} = \hat{\mathbf{Y}}_{k,cube}[n^x_R,n^z_R,n,m] \!\!= \\ &\!  \mathscr{G}_{k}  
e^{\!-\!j4\pi\! f_m \!  \frac{r_{k}}{c}}\!
e^{j4\pi \!f_0\!\frac{v_{r,k}\! n\!T_s}{c}}\!\!
e^{j\!\frac{\pi\! f_0d(\Omega_{k,n}\!-\Omega_{k}\!)(N_H^z\!-\!1)}{c}}\!\!
e^{j\!\frac{\pi\! f_0d(\Psi_{k,n}\!\!-\Psi_{k})(N_H^x\!-\!1)\!}{c}}\!\!
e^{j\!\frac{2\pi\! f_0n_R^zd\Omega_{k,n}}{c}}\!\! 
e^{j\!\frac{2\pi\! f_0n_R^xd\Psi_{k,n}}{c}}
\!\!+\!  \hat{n}_{k,n^x_R\!,n^z_R,n,m}.
			\end{aligned}
		\end{split}\tag{46}
	\end{equation}
	\setcounter{equation}{\value{MYtempeqncnt}}
	\hrulefill 
	\vspace*{8pt} 
\end{figure*}

\section{6D Motion  Parameters Estimation   Scheme}

In this section, we analyze the sensing echo signals of ISAC system,
and   propose a novel 6D motion  parameters estimation   scheme for dynamic target sensing.

\subsection{Collection of Sensing Echo Signals }

The premise for ISAC system to sense the $k$-th dynamic target is that the transmitting beam of HU-UPA illuminates the $k$-th target. In  SBS stage,  
the sensing beam of ISAC system can explore all angle spaces at certain scanning intervals. Thus, within a certain UTS, the BS will  illuminate the $k$-th target. 
In  SBT stage, the ISAC system can continuously predict and track the $k$-th target using the algorithms such as Kalman filtering\cite{Ma2020}, thereby illuminating the $k$-th target with the transmitted tracking beam.
Nevertheless, since our focus hre is to estimate the motion parameters of the $k$-th dynamic target, we will directly assume that  HU-UPA of  BS  generates one sensing beam towards $(\theta_{k}, \phi_{k})$ during one UTS to sense the $k$-th target. 

Based on (8), the  transmission signals of HU-UPA during one UTS can be expressed  as 
\begin{equation}
	\begin{split}
		\begin{aligned}
			\label{deqn_ex1a}
&  \!\!\grave{\mathbf{x}}_{\theta_{k}, \phi_{k},n,m} \!=\!  \sqrt{\!\frac{\grave{\rho}_{\theta_{k}, \phi_{k}} \!P_t}{N_H}}\mathbf{a}_{H}\left(\Psi({\theta_{k}, \phi_{k}}),\Omega(\theta_{k}, \phi_{k})\right)s^{\theta_{k}, \phi_{k}}_{n,m}
\\&\!\!  \!\!=\sqrt{\!\frac{\grave{\rho}_{\theta_{k}, \phi_{k}} \!P_t}{N_H}}\mathbf{a}_{H}(\cos \phi_{k} \cos \theta_{k},\sin \phi_{k})s^{\theta_{k}, \phi_{k}}_{n,m}.
		\end{aligned}
	\end{split}\tag{38}
\end{equation}
Then the sensing echo signals on the $m$-th subcarrier of the $n$-th OFDM symbol received by  RU-UPA can be represented as
\begin{equation}
	\begin{split}
		\begin{aligned}
			\label{deqn_ex1a}
&\mathbf{y}_{k,n,m} \!=\!  \mathbf{H}^{sensing}_{n,m}  \grave{\mathbf{x}}_{\theta_{k}, \phi_{k},n,m}^* \!+\! \mathbf{n}_{k,n,m} \\& =  \mathbf{H}^{target}_{n,m} \grave{\mathbf{x}}_{\theta_{k}, \phi_{k},n,m}^* \!+\! \mathbf{H}^{background}_{n,m} \grave{\mathbf{x}}_{\theta_{k}, \phi_{k},n,m}^* \!+\! \mathbf{n}_{k,n,m}, 
\\&\kern 55pt  n=0,...,N-1,\kern 4pt m=0,...,M-1,
		\end{aligned}
	\end{split}\tag{39}
\end{equation}
where the element of $\mathbf{n}_{k,n,m}$ is  zero-mean additive   Gaussian   noise  with variance  $\sigma_{k}^2$.
Since $\mathbf{y}_{k,n,m}$ represents the echo signals received by all $N_R=N_R^x\times N_R^z$ receiving antennas, we  reformat   vector $\mathbf{y}_{k,n,m}$ into matrix form as
\begin{equation}
	\begin{split}
		\begin{aligned}
			\label{deqn_ex1a}
\mathbf{Y}_{k,n,m} =  {\rm reshape} \{\mathbf{y}_{k,n,m},[N_R^x,N_R^z]\}
\in \mathbb{C}^{N_{R}^x\times N_{R}^z}.
		\end{aligned}
	\end{split}\tag{40}
\end{equation}
Moreover, we can stack $\mathbf{Y}_{k,n,m}$ of 
all $N$ symbols and  all $M$ subcarriers into one echoes tensor 
${\mathbf{Y}}_{k,cube} \in \mathbb{C}^{N_{R}^x\times N_{R}^z \times N\times M}$, 
whose $(n^x_R,n^z_R,n,m)$-th element is ${\mathbf{Y}}_{k,cube}[n^x_R,n^z_R,n,m] = [\mathbf{Y}_{k,n,m}]_{n^x_R,n^z_R}$.

Note that ${\mathbf{Y}}_{k,cube}$ includes the sensing  channel $\mathbf{H}^{sensing}_{n,m}$, transmitting beamforming, and transmission symbols $s^{\theta_{k}, \phi_{k}}_{n,m}$, by which means the  targets sensing can be understood as an estimation of $\mathbf{H}^{sensing}_{n,m}$. However, random transmission symbols would affect the estimation of  sensing  channel, and thus we need to  erase the transmission symbols from the received signals to obtain  equivalent echo  channel (EEC).
Specifically, the EEC corresponding to $\mathbf{Y}_{k,n,m}$ can be obtained as
$\tilde{\mathbf{Y}}_{k,n,m} = \mathbf{Y}_{k,n,m}/s^{\theta_{k}, \phi_{k}}_{n,m}$.
Then we can stack $\tilde{\mathbf{Y}}_{k,n,m}$ 
into an EEC tensor $\tilde{\mathbf{Y}}_{k,cube} \in \mathbb{C}^{N_{R}^x\times N_{R}^z \times N\times M}$
with $\tilde{\mathbf{Y}}_{k,cube}[n^x_R,n^z_R,n,m] = [\tilde{\mathbf{Y}}_{k,n,m}]_{n^x_R,n^z_R}$.

\newcounter{TempEqCnt} 
\setcounter{TempEqCnt}{\value{equation}} 
\setcounter{equation}{4} 
\begin{figure*}[ht] 
	\vspace{-8mm}
\begin{equation}
	\begin{split}
		\begin{aligned}
			\label{deqn_ex1a}
\Psi_{k,n} &= \cos (\phi_{k} - \omega_{\phi,k}nT_{s})\cos (\theta_{k}-\omega_{\theta,k}nT_{s})
 \\& = \cos \phi_{k}\cos \theta_{k} + 
\sin\phi_{k}\cos\theta_{k}\omega_{\phi,k}nT_s+ \cos\phi_{k}\sin\theta_{k}
\omega_{\theta,k}nT_s\\&=
\Psi_{k} + 
\sin\phi_{k}\cos\theta_{k}\omega_{\phi,k}nT_s+ \cos\phi_{k}\sin\theta_{k}
\omega_{\theta,k}nT_s.
		\end{aligned}
	\end{split}\tag{48}
\end{equation}
	\begin{equation}
		\begin{split}
			\begin{aligned}
				\label{deqn_ex1a}
\hat{y}_{k,n^x_R,n^z_R,n,m} = &\mathscr{G}_{k}
\underbrace{e^{\!-\!j4\pi\! f_m \!\!  \frac{r_{k}}{c}}\!\!}_{\rm distance} \underbrace{
e^{j\!\frac{2\pi\! f_0d}{c}\!\sin\!\phi_{k}n^z_R}\!}_{\rm pitch \kern 2pt angle} \underbrace{e^{j\!\frac{2\pi\! f_0d}{c}\!\cos\phi_{k}\!\cos\!\theta_{k}n^x_R}}_{\rm horizontal \kern 2pt angle}
\times  \underbrace{e^{j4\pi \!f_0\!\frac{v_{r,k}n\!T_s}{c}}}_{\rm radial \kern 2pt velocity} \times \\&
\underbrace{e^{\!-j\!\frac{\pi\! f_0d}{c}[(N_H^z\!-\!1)\!\cos\!\phi_{k}-(\!N^x_H\!-\!1)\!\sin\!\phi_{k}\!\cos\!\theta_{k}]\omega_{\phi,k}nT_s}e^{\!-j\!\frac{2\pi f_0d}{c}\!(n_R^z\cos\phi_{k}-n^x_R\sin\phi_{k}\cos\theta_{k})\omega_{\phi,k}nT_s}}_{\rm pitch \kern 2pt angular \kern 2pt velocity} \times \\&
\underbrace{e^{j\frac{\pi f_0d(N_H^x-1)}{c}\cos\phi_{k}\sin\theta_{k}\omega_{\theta,k}nT_s}
e^{j\frac{2\pi f_0 n_R^xd}{c}\cos\phi_{k}\sin\theta_{k}\omega_{\theta,k}nT_s}}_{\rm horizontal \kern 2pt angular \kern 2pt velocity}+ \hat{n}_{k,n^x_R,n^z_R,n,m}
			\end{aligned}
		\end{split}\tag{49}
	\end{equation}
	\hrulefill 
	\vspace*{8pt} 
\end{figure*}

\subsection{Filtering the Static Environmental Clutter}

It is known from (39) and (40) that
the echo signal ${\mathbf{Y}}_{k,cube}$ includes both dynamic target echoes and static environment echoes, 
while the  EEC $\tilde{\mathbf{Y}}_{k,cube}$  also includes both the
EEC of  dynamic targets (DT-EEC) and the
EEC of  static environment (SE-EEC).
The SE-EEC in  original echo signals would cause negative interference to dynamic target sensing, and thus  SE-EEC can be referred to as  clutter-EEC.
To address this negative interference,  we need  to filter out the interference of
clutter-EEC  and  extract   effective DT-EEC from $\tilde{\mathbf{Y}}_{k,cube}$.
According to  clutter suppression method   \cite{10477890}, we may express the effective DT-EEC after static clutter filtering as $\hat{\mathbf{Y}}_{k,cube}$,
whose $[:,:,n,m]$-th sub-matrix is
$\hat{\mathbf{Y}}_{k,cube}[:,:,n,m]=
\hat{\mathbf{Y}}_{k,n,m} = {\rm reshape} \{\hat{\mathbf{y}}_{k,n,m},[N_R^x,N_R^z]\}$
with  
\begin{equation}
	\begin{split}
		\begin{aligned}
			\label{deqn_ex1a}
 \hat{\mathbf{y}}_{k,n,m}  & \approx   \mathbf{H}^{target}_{n,m} \grave{\mathbf{x}}_{\theta_{k}, \phi_{k},n,m}^*/s^{\theta_{k}, \phi_{k}}_{n,m} \!+\!  \hat{\mathbf{n}}_{k,n,m}
\\& = \sum_{k'=1}^K \mathbf{H}_{k',n,m} \grave{\mathbf{x}}_{\theta_{k}, \phi_{k},n,m}^*/s^{\theta_{k}, \phi_{k}}_{n,m} \!+\!  \hat{\mathbf{n}}_{k,n,m}, 
		\end{aligned}
	\end{split}\tag{41}
\end{equation}
where $\hat{\mathbf{n}}_{k,n,m}$  is the noise after static clutter filtering.

\subsection{Sensing Echo Signals Analysis}

Based on (29), (34) and (38),  
$\hat{\mathbf{y}}_{k,n,m}$ in (41) can be calculated as (42) at the top of the this page. 
Let us record 
\begin{align}
	\Psi_{k}&= \Psi({\theta_{k}, \phi_{k}}) = \cos \phi_{k} \cos \theta_{k}, \tag{43}\\
	\Omega_{k}&=\Omega(\theta_{k}, \phi_{k})= \sin \phi_{k}.
	\tag{44}
\end{align} 

Since $K$ dynamic targets have different directions, (42) can be calculated as (45) at the top of  this page,
where $\mathscr{G}_{k} = \alpha_{k} \sqrt{\frac{\grave{\rho}_{\theta_{k}, \phi_{k}}\!P_t}{N_H}}
\!\frac{\sin [\!\frac{\pi\!f_0d}{c} (\Omega_{k\!,n}\!-\Omega_{k})\!N_H^z\!]}
{\sin [\frac{\pi\!f_0d}{c} (\Omega_{k\!,n}\!-\Omega_{k})]}
\!\frac{\sin [\!\frac{\pi\!f_0d}{c} (\Psi_{k\!,n}\!-\Psi_{k})\!N_H^x\!]}
{\sin [\frac{\pi\!f_0d}{c} (\Psi_{k\!,n}\!-\Psi_{k})]}$.
Then, based on   (45),
the $[n^x_R,n^z_R,n,m]$-th element in 
$\hat{\mathbf{Y}}_{k,cube}$ can be 
expressed as (46) at the top of  this page.

To further simplify (46), we note that $f(a+h)\approx f(a)+f'(a)h$.
Thus there is 
\begin{equation}
	\begin{split}
		\begin{aligned}
			\label{deqn_ex1a}
\!\!\!\Omega_{k,n} &= \sin(\phi_{k}-\omega_{\phi,k}nT_s) \approx \sin \phi_{k} \!-\! \cos \phi_{k} \omega_{\phi,k}nT_s
\\&=\Omega_{k}-\cos \phi_{k} \omega_{\phi,k}nT_s.
		\end{aligned}
	\end{split}\tag{47}
\end{equation}
From the approximation $f(a+h,b+u)\approx f(a,b)+f'_a(a,b)h+f'_b(a,b)u$, $\Psi_{k,n}$ can be computed 
as shown in (48) at the top of  the next page.

Based on (47) and (48),
the $[n^x_R,n^z_R,n,m]$-th element in 
$\hat{\mathbf{Y}}_{k,cube}$, i.e., 
the $\hat{y}_{k,n^x_R,n^z_R,n,m}$  in (46) 
can be  calculated as shown in (49) at the top of  the next page.
Eq. (49) indicates that $\hat{y}_{k,n^x_R,n^z_R,n,m}$ includes the Dis, PA, HA, RV, PAV, and HAV, and therefore, we could estimate the 6D motion parameters of dynamic targets from DT-EEC $\hat{\mathbf{Y}}_{k,cube}$.

\subsection{Estimation of HA, PA and Dis}

Let us transform $\hat{\mathbf{Y}}_{k,cube} \in \mathbb{C}^{N_{R}^x\times N_{R}^z \times N\times M}$ into an $\Omega$-matrix $\mathbf{Y}_{k,\Omega}$ with  dimension $N_R^z\times N_R^xNM$. Based on (49), $\mathbf{Y}_{k,\Omega}$  can be represented as
\begin{equation}
	\begin{split}
		\begin{aligned}
			\label{deqn_ex1a}
\mathbf{Y}_{k,\Omega}= \mathbf{k}_{\Omega}(\Omega_{k})\cdot
\mathbf{x}_{k,\Omega} + \mathbf{N}_{k,\Omega} \in \mathbb{C}^{N_R^z\times N_{R}^x  N M},
		\end{aligned}
	\end{split}\tag{50}
\end{equation}
where $\Omega_{k} = \sin\phi_{k}$, $\mathbf{x}_{k,\Omega}\in \mathbb{C}^{1\times N_{R}^x  N M}$, $\mathbf{N}_{k,\Omega} \in \mathbb{C}^{N_R^z\times N_{R}^x  N M}$, and $\mathbf{k}_{\Omega}(\Omega) = [1,e^{j\frac{2\pi f_0d\Omega}{c}},...,e^{j\frac{2\pi f_0d\Omega}{c}(N_{R}^z-1)}]^T \!\! \in\! \mathbb{C}^{N_{R}^z\times 1}$ is defined as the \emph{pitch spatial-domain direction array steering vector}.  
Since $\mathbf{Y}_{k,\Omega}$ is the array signal related to the pitch spatial-domain direction array, we can estimate $\Omega_{k}$ from $\mathbf{Y}_{k,\Omega}$ by utilizing array signal processing methods.

Here we adopt the
estimating signal
parameters via rotational variation techniques
(ESPRIT) method for parameter estimation\cite{32276,1164935,671426}.
Specifically, the covariance matrix of $\mathbf{Y}_{k,\Omega}$ can be calculated as
$\mathbf{R}_{k,\Omega} = \frac{1}{N^x_RNM}\mathbf{Y}_{k,\Omega}(\mathbf{Y}_{k,\Omega})^H$. 
We perform eigenvalue decomposition of  $\mathbf{R}_{k,\Omega}$  to obtain the diagonal matrix
$\mathbf{\Sigma}_{k,\Omega}$ and the corresponding eigenvector matrix $\mathbf{U}_{k,\Omega}$, that is, $[\mathbf{U}_{k,\Omega}, \mathbf{\Sigma}_{k,\Omega}]={\rm eig}(\mathbf{R}_{k,\Omega})$.
Then  the minimum description length (MDL) criterion is utilized to estimate the number of dynamic targets from $\mathbf{\Sigma}_{k,\Omega}$  as $K_{k,\Omega}$\cite{mdl,mdl2}. We  extract the  parallel signal subspaces from $\mathbf{U}_{k,\Omega}$ as 
$\mathbf{U}_{k,\Omega,1} = \left[ \mathbf{U}_{k,\Omega}[1:N_R^z-1,1:K_{k,\Omega}], \mathbf{U}_{k,\Omega}[2:N_R^z,1:K_{k,\Omega}] \right] \in \mathbb{C}^{(N_R^z-1)\times 2K_{k,\Omega}}$ and compute 
$\tilde{\mathbf{R}}_{k,\Omega} = (\mathbf{U}_{k,\Omega,1})^H\mathbf{U}_{k,\Omega,1}\in\mathbb{C}^{2K_{k,\Omega}\times 2K_{k,\Omega}}$. Then we perform eigenvalue decomposition of 
$\tilde{\mathbf{R}}_{k,\Omega}$ to obtain the diagonal matrix 
$\tilde{\mathbf{\Sigma}}_{k,\Omega}$ and the corresponding eigenvector matrix $\tilde{\mathbf{U}}_{k,\Omega}$, that is, $[\tilde{\mathbf{U}}_{k,\Omega}, \tilde{\mathbf{\Sigma}}_{k,\Omega}]={\rm eig}(\tilde{\mathbf{R}}_{k,\Omega})$.
We extract $\tilde{\mathbf{U}}_{k,\Omega,a} = \tilde{\mathbf{U}}_{k,\Omega}[1:K_{k,\Omega},K_{k,\Omega}+1:2K_{k,\Omega}] \in \mathbb{C}^{K_{k,\Omega}\times K_{k,\Omega}}$ and 
$\tilde{\mathbf{U}}_{k,\Omega,b} = \tilde{\mathbf{U}}_{k,\Omega}[K_{k,\Omega}+1:2K_{k,\Omega},K_{k,\Omega}+1:2K_{k,\Omega}] \in \mathbb{C}^{K_{k,\Omega}\times K_{k,\Omega}}$. Then we can compute $\check{\mathbf{R}}_{k,\Omega}=-\tilde{\mathbf{U}}_{k,\Omega,a}(\tilde{\mathbf{U}}_{k,\Omega,b})^{-1}$. Next, we perform eigenvalue decomposition of  $\check{\mathbf{R}}_{k,\Omega}$  to obtain the diagonal matrix 
$\check{\mathbf{\Sigma}}_{k,\Omega}$
and the corresponding eigenvector matrix $\check{\mathbf{U}}_{k,\Omega}$, that is, $[\check{\mathbf{U}}_{k,\Omega}, \check{\mathbf{\Sigma}}_{k,\Omega}]={\rm eig}(\check{\mathbf{R}}_{k,\Omega})$.
We take out the elements on the main diagonal of $\check{\mathbf{\Sigma}}_{k,\Omega}$ to form one  eigenvalues set as $\{\lambda_{k,\Omega,1},\lambda_{k,\Omega,2},...,\lambda_{k,\Omega,K_{k,\Omega}}\}$,
and  compute the \emph{space values} as
$\kappa_{k,\Omega,i} =\arctan{\frac{\mathcal{I}(\lambda_{k,\Omega,i})}{\mathcal{R}(\lambda_{k,\Omega,i})}}$, where 
$i=1,2,...,K_{k,\Omega}$.
Since $\hat{\mathbf{Y}}_{k,cube}$ only contains one dynamic target, there should be 
$K_{k,\Omega}=1$, and thus we  abbreviate the space value corresponding to $\mathbf{Y}_{k,\Omega}$ as $\kappa_{k,\Omega}$.
 Then the pitch spatial-domain direction of the $k$-th dynamic target can be estimated as 
\begin{equation}
	\begin{split}
		\begin{aligned}
			\label{deqn_ex1a}
\hat{\Omega}_{k} =\frac{c\kappa_{k,\Omega}}{2\pi f_0 d}.
		\end{aligned}
	\end{split}\tag{51}
\end{equation}
Finally, the PA of the $k$-th dynamic target  can be estimated as 
\begin{equation}
	\begin{split}
		\begin{aligned}
			\label{deqn_ex1a}
\hat{\phi}_{k} = \arcsin \left(		\hat{\Omega}_{k}\right)	 =\arcsin \left(	\frac{c\kappa_{k,\Omega}}{2\pi f_0 d}\right).
		\end{aligned}
	\end{split}\tag{52}
\end{equation}

\setcounter{TempEqCnt}{\value{equation}} 
\setcounter{equation}{4} 
\begin{figure*}[ht] 
	\begin{equation}
		\begin{split}
			\begin{aligned}
				\label{deqn_ex1a}
&v^{vir}_{k,n^x_R,n^z_R} = v_{r,k}	-\frac{d}{4}\left[(N_H^z-1)\cos\phi_{k}
-(N_H^x-1)\sin\phi_{k}\cos\theta_{k}\right]\omega_{\phi,k} -\!
\frac{d}{2}(n_R^z\cos\phi_{k}\!-\!n^x_R\sin\phi_{k}\cos\theta_{k})\omega_{\phi,k}\!\\&\quad \quad \quad\quad\quad+\!
\frac{d}{4}(N^x_H\!-\!1)\cos\phi_{k}\sin\theta_{k}\omega_{\theta,k}
\!+\!\frac{d}{2}n^x_R\cos\phi_{k}\sin\theta_{k}\omega_{\theta,k}.
			\end{aligned}
		\end{split}\tag{58}
	\end{equation}
	\begin{equation}
		\begin{split}
			\begin{aligned}
				\label{deqn_ex1a}
\hat{y}_{k,n^x_R,n^z_R,n,m} = &\mathscr{G}_{k} 
\underbrace{e^{\!-\!j4\pi\! f_m \!\!  \frac{r_{k}}{c}}\!\!}_{\rm distance} \underbrace{
e^{j\!\frac{2\pi\! f_0d}{c}\!\sin\!\phi_{k}n^z_R}\!}_{\rm pitch \kern 2pt angle} \underbrace{e^{j\!\frac{2\pi\! f_0d}{c}\!\cos\phi_{k}\!\cos\!\theta_{k}n^x_R}}_{\rm horizontal \kern 2pt angle} \underbrace{e^{j4\pi \!f_0\!\frac{v^{vir}_{k,n^x_R,n^z_R}n\!T_s}{c}}}_{\rm virtual-velocity} 
+ \hat{n}_{k,n^x_R,n^z_R,n,m}
			\end{aligned}
		\end{split}\tag{59}
	\end{equation}
	\hrulefill 
	\vspace*{8pt} 
\end{figure*}

Similarly, let us transform $\hat{\mathbf{Y}}_{k,cube} \in \mathbb{C}^{N_{R}^x\times N_{R}^z \times N\times M}$ into an $\Psi$-matrix $\mathbf{Y}_{k,\Psi}$ with  dimension  $N_R^x\times N_R^zNM$. Based on (49), $\mathbf{Y}_{k,\Psi}$  can be represented as
\begin{equation}
	\begin{split}
		\begin{aligned}
			\label{deqn_ex1a}
			\mathbf{Y}_{k,\Psi}= \mathbf{k}_{\Psi}(\Psi_{k})\cdot
			\mathbf{x}_{k,\Psi} + \mathbf{N}_{k,\Psi} \in \mathbb{C}^{N_R^x\times N_{R}^z  N M},
		\end{aligned}
	\end{split}\tag{53}
\end{equation}
where $\Psi_{k} = \cos\phi_{k}\cos\theta_{k}$, $\mathbf{x}_{k,\Psi}\in \mathbb{C}^{1\times N_{R}^z  N M}$, $\mathbf{N}_{k,\Psi} \in \mathbb{C}^{N_R^x\times N_{R}^z  N M}$, and $\mathbf{k}_{\Psi}(\Psi) = [1,e^{j\frac{2\pi f_0d\Psi}{c}},...,e^{j\frac{2\pi f_0d\Psi}{c}(N_{R}^x-1)}]^T \!\! \in\! \mathbb{C}^{N_{R}^x\times 1}$ is defined as the \emph{horizontal spatial-domain direction array steering vector}.
Similarly, we can employ the ESPRIT method to obtain the space value  corresponding to $\mathbf{Y}_{k,\Psi}$ as $\kappa_{k,\Psi}$.
Then  the horizontal spatial-domain direction of the $k$-th dynamic target  can be estimated as 
\begin{equation}
	\begin{split}
		\begin{aligned}
			\label{deqn_ex1a}
			\hat{\Psi}_{k} =\frac{c\kappa_{k,\Psi}}{2\pi f_0 d},
		\end{aligned}
	\end{split}\tag{54}
\end{equation}
and  the HA  of the $k$-th dynamic target   can be estimated as 
\begin{equation}
	\begin{split}
		\begin{aligned}
			\label{deqn_ex1a}
\hat{\theta}_{k} = \arccos \left(		\frac{\hat{\Psi}_{k}}{\cos \hat{\phi}_{k}}  \right).
		\end{aligned}
	\end{split}\tag{55}
\end{equation}

To estimate the Dis of the target,
let us   transform $\hat{\mathbf{Y}}_{k,cube} \in \mathbb{C}^{N_{R}^x\times N_{R}^z \times N\times M}$ into a distance-matrix $\mathbf{Y}_{k,r}$ with  dimension  $M\times N_R^x N_R^zN$. Based on (49), $\mathbf{Y}_{k,r}$  can be represented as
\begin{equation}
	\begin{split}
		\begin{aligned}
			\label{deqn_ex1a}
			\mathbf{Y}_{k,r}= \mathbf{k}_{r}(r_{k})\cdot
			\mathbf{x}_{k,r} + \mathbf{N}_{k,r} \in \mathbb{C}^{M\times N_R^x N_R^zN},
		\end{aligned}
	\end{split}\tag{56}
\end{equation}
where  $\mathbf{x}_{k,r}\in \mathbb{C}^{1\times N_R^x N_R^zN}$, $\mathbf{N}_{k,r} \in \mathbb{C}^{M\times N_R^xN_R^zN}$, and $\mathbf{k}_{r}(r) = [1,e^{-j\frac{4\pi r\Delta f}{c}},...,e^{-j\frac{4\pi r\Delta f}{c}(M-1)}]^T \in \mathbb{C}^{M\times 1}$ is defined as the \emph{distance array steering vector}.
Similarly, we can employ the ESPRIT method to obtain the space value  corresponding to $\mathbf{Y}_{k,r}$ as $\kappa_{k,r}$.
Then the Dis of the $k$-th dynamic target  can be estimated as 
\begin{equation}
	\begin{split}
		\begin{aligned}
			\label{deqn_ex1a}
			\hat{r}_{k} =-\frac{c\kappa_{k,r}}{4\pi \Delta f}.
		\end{aligned}
	\end{split}\tag{57}
\end{equation}

\subsection{Estimation of RV, HAV  and PAV}

It can be analyzed from (49) that each antenna observes one \emph{virtual-velocity} composed of  RV, HAV, and PAV of the  target. Note that the virtual-velocity  observed by different antenna is different. Thus, we can  derive the virtual-velocity of the $k$-th dynamic target observed by the $(n^x_R,n^z_R)$-th antenna  from (49) as $v^{vir}_{k,n^x_R,n^z_R}$, which is shown in (58) at the top of this page.
Then, (49) can be rewritten as (59) at the top of this page.

Next, we need to estimate the virtual-velocity observed by each antenna.
Let us extract DT-EEC of the $(n^x_R,n^z_R)$-th antenna on all subcarriers of all OFDM symbols from $\hat{\mathbf{Y}}_{k,cube} \in \mathbb{C}^{N_{R}^x\times N_{R}^z \times N\times M}$ as
\begin{equation}
	\begin{split}
		\begin{aligned}
			\label{deqn_ex1a}
&\mathbf{Y}_{k,v_{vir},n^x_R,n^z_R}\!\\&=\! \mathbf{k}_{v_{vir}}(v^{vir}_{k,n^x_R,n^z_R})\!\cdot\!
\mathbf{x}_{k,v_{vir},n^x_R,n^z_R} \!+\! \mathbf{N}_{k,v_{vir},n^x_R,n^z_R},
		\end{aligned}
	\end{split}\tag{60}
\end{equation}
where $\mathbf{Y}_{k,v_{vir},n^x_R,n^z_R} \in \mathbb{C}^{N\times M}$, $\mathbf{x}_{k,v_{vir},n^x_R,n^z_R}\in \mathbb{C}^{1\times M}$, $\mathbf{N}_{k,v_{vir},n^x_R,n^z_R} \in \mathbb{C}^{N\times M}$,  $[\mathbf{Y}_{k,v_{vir},n^x_R,n^z_R}]_{n,m}=\hat{y}_{k,n^x_R,n^z_R,n,m}$, and 
$\mathbf{k}_{v_{vir}}(v_{vir}) = [1,e^{j\frac{4\pi f_0v_{vir}T_s}{c}},...,e^{j\frac{4\pi f_0v_{vir}T_s}{c}(N-1)}]^T \in \mathbb{C}^{N\times 1}$ is defined as the \emph{virtual-velocity array steering vector}. 
Similarly, we can employ the ESPRIT method to obtain the space value  corresponding to $\mathbf{Y}_{k,v_{vir},n^x_R,n^z_R}$ as $\kappa_{k,v_{vir},n^x_R,n^z_R}$.
Then the virtual-velocity of the $k$-th dynamic target observed by the $(n^x_R,n^z_R)$-th antenna can be estimated as 
\begin{equation}
	\begin{split}
		\begin{aligned}
			\label{deqn_ex1a}
\hat{v}^{vir}_{k,n^x_R,n^z_R} =\frac{c \kappa_{k,v_{vir},n^x_R,n^z_R}}{4\pi f_0T_s}.
		\end{aligned}
	\end{split}\tag{61}
\end{equation}
By traversing each antenna, we can obtain the virtual-velocity observed by each antenna, 
which is record as 
$(n_R^x,n_R^z,\hat{v}^{vir}_{k,n^x_R,n^z_R})$ with  $n_{R}^x \in \{0,1,...,N^x_{R}-1\}$ and $n_{R}^z  \in \{0,1,...,N^z_{R}-1\}$.
Next we need to estimate the RV, HAV, and PAV of the target from these $N_R=N_R^xN_R^z$ ternary pairs.

In fact, we can express  $v^{vir}_{k,n^x_R,n^z_R}$ as a binary function of $(n_R^x,n_R^z)$. Based on (58),
there is
\begin{equation}
	\begin{split}
		\begin{aligned}
			\label{deqn_ex1a}
{v}^{vir}_{k,n^x_R,n^z_R} = A_{k}+B_{k}\cdot n^x_R+C_{k} \cdot n^z_R,
		\end{aligned}
	\end{split}\tag{62}
\end{equation}
where
\begin{equation}
	\begin{split}
		\begin{aligned}
			\label{deqn_ex1a}
A_{k}\!=&  v_{r\!,k} \!-\!\frac{d}{4}\!\left[\!(\!N_H^z\!\!-\!\!1)\!\cos\!\phi_{k}
\!\!-\!\!(\!N_H^x\!\!-\!\!1\!)\!\sin\!\phi_{k}\!\cos\!\theta_{k}\!\right]\!\omega_{\phi,k}
\\& +\frac{d}{4}
(N^x_H\!-\!1)\cos\phi_{k}\sin\theta_{k}\omega_{\theta,k},
		\end{aligned}
	\end{split}\tag{63}
\end{equation}
\begin{equation}
	\begin{split}
		\begin{aligned}
			\label{deqn_ex1a}
\!\!\!\!\!\!
\!\!\!\!\!\!\!\!\!\!\!
B_{k}\!=&  \frac{d}{2} (\sin\!\phi_{k}\cos\!\theta_{k}\omega_{\phi,k}\!+\!
\cos\phi_{k}\sin\theta_{k}\omega_{\theta,k}),
		\end{aligned}
	\end{split}\tag{64}
\end{equation}
\begin{equation}
	\begin{split}
		\begin{aligned}
			\label{deqn_ex1a}
\!\!\!\!\!\!\!\!\!\!\!\!\!\!\!
\!\!\!\!\!\!\!\!\!\!\!\!\!\!\!\!\!\!\!\!
\!\!\!\!\!\!\!\!\!\!\!\!\!\!\!\!\!\!\!\!
\!\!\!\!\!\!\!\!\!\!\!
C_{k}\!=&  -\frac{d}{2} \cos\phi_{k} \omega_{\phi,k}.
		\end{aligned}
	\end{split}\tag{65}
\end{equation}
Eq. (62) indicates that  the ternary pairs $(n_R^x,n_R^z,{v}^{vir}_{k,n^x_R,n^z_R})$ could form a plane in three-dimensional space.

\begin{figure}[!t]
	\centering
	\includegraphics[width=85mm]{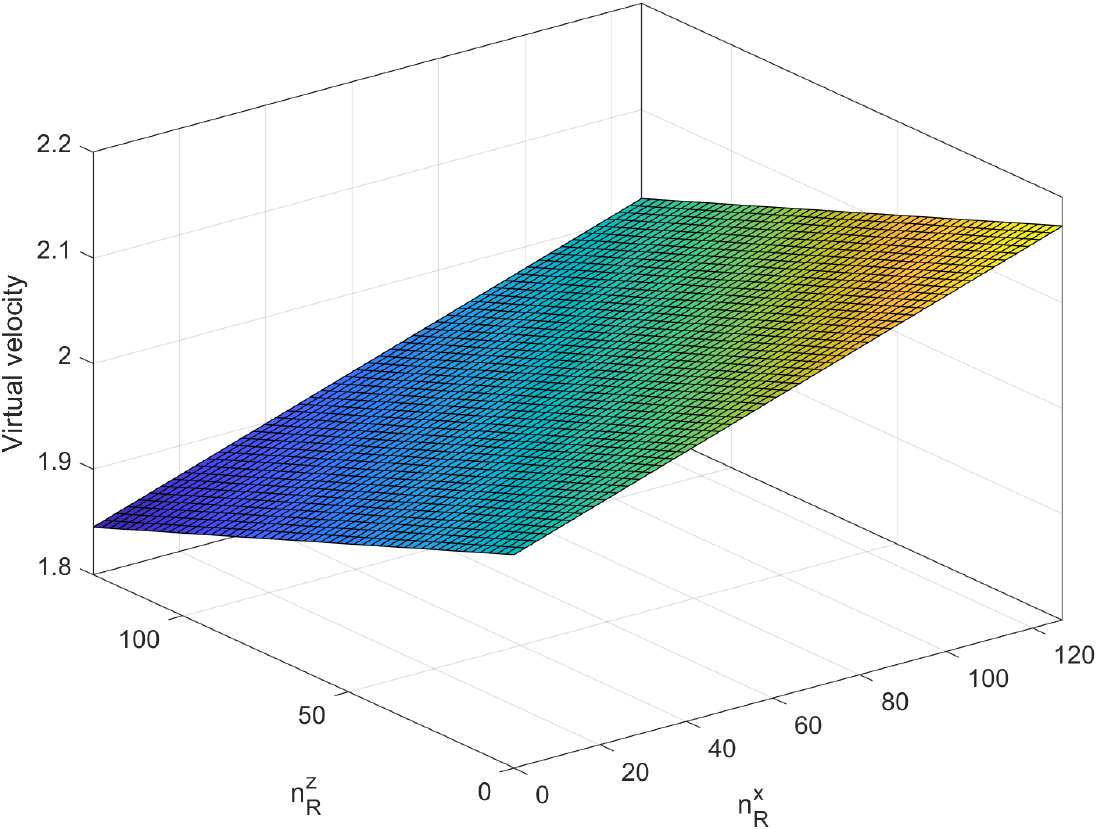}
	\caption{Virtual velocities observed by different antennas.}
	\label{fig_1}
\end{figure}

Therefore, we can   use the least squares (LS) method to
planar fit the $\{(n_R^x,n_R^z,\hat{v}^{vir}_{k,n^x_R,n^z_R})|n_{R}^x =0,1,...,N^x_{R}-1; n_{R}^z =0,1,...,N^z_{R}-1\}$, and  record the parameter results of plane fitting as
$\hat{A}_{k}$, $\hat{B}_{k}$ and
$\hat{C}_{k}$. 

Based on (63), (64) and (65),
the PAV, HAV, and RV of the $k$-th dynamic target  can be sequentially estimated as
\begin{equation}
	\begin{split}
		\begin{aligned}
			\label{deqn_ex1a}
\!\!\!\!\!
\!\!\!\!\!\!\!\!\!\!\!\!\!\!\!\!\!\!\!\!\!\!\!\!\!\!\!\!\!\!\!\!\!\!\!\!\!\!\!\!\!\!\!\!\!\!\!\!\!\!\!\!\!\!\!\!\!\!\!\!\!\!\!\!
\hat{\omega}_{\phi,k}=&  -\frac{2\hat{C}_{k,l}}{d\cos\hat{\phi}_{k}},
		\end{aligned}
	\end{split}\tag{66}
\end{equation}
\begin{equation}
	\begin{split}
		\begin{aligned}
			\label{deqn_ex1a}
\!\!\!\!\!\!
\!\!\!\!\!\!\!\!\!\!\!
\!\!\!\!\!\!\!\!\!\!\!\!\!\!\!\!\!\!\!\!
\hat{\omega}_{\theta,k}\!=&  -\frac{2\hat{B}_{k}/d -\sin\hat{\phi}_{k}\cos\hat{\theta}_{k} \hat{\omega}_{\phi,k}}{\cos\hat{\phi}_{k}\sin\hat{\theta}_{k}},
		\end{aligned}
	\end{split}\tag{67}
\end{equation}
\begin{equation}
	\begin{split}
		\begin{aligned}
			\label{deqn_ex1a}
\!\!\!\!&\hat{v}_{r,k}\!=\!\! \hat{A}_{k}\!+\!
\!\frac{d}{4}\!\left[\!(\!N_{\!H}^z\!\!-\!\!1\!)\!\cos\!\hat{\phi}_{k}
\!\!-\!\!(\!N_{\!H}^x\!-\!\!1\!)\!\sin\!\hat{\phi}_{k}\!\cos\!\hat{\theta}_{k}\!\right]\!\!\hat{\omega}_{\phi,k} \\&         	
\kern 35pt -\frac{d}{4}
(N^x_H\!-\!1)\cos\hat{\phi}_{k}\sin\hat{\theta}_{k}\hat{\omega}_{\theta,k}.
		\end{aligned}
	\end{split}\tag{68}
\end{equation}

Clearly, from  (52), (55), (57), (66), (67) and (68) 
we have completed the estimation of the 6D motion parameters of the $k$-th  dynamic target. 
The proposed 6D motion  parameters estimation scheme is summarized in Algorithm 1, as shown on the next page.

\begin{algorithm}[!t]
	\caption{6D Motion  Parameters Estimation   Scheme}
	\label{alg2}
	\begin{algorithmic}[1]
		\REQUIRE ${\mathbf{Y}}_{k,cube} \in \mathbb{C}^{N_{R}^x\times N_{R}^z \times N\times M}$
		\ENSURE $\hat{r}_{k}, \hat{\theta}_{k}, \hat{\phi}_{k}, \hat{v}_{r,k},\hat{\omega}_{\theta,k},\hat{\omega}_{\phi,k}$
		\STATE Perform symbol erasure on ${\mathbf{Y}}_{k,cube}$ to obtain $\tilde{\mathbf{Y}}_{k,n,m}$.
\STATE Perform clutter suppression on $\tilde{\mathbf{Y}}_{k,n,m}$ to obtain $\hat{\mathbf{Y}}_{k,cube}$.
\STATE Transform $\hat{\mathbf{Y}}_{k,cube}$ into $\mathbf{Y}_{k,\Omega}$, and then perform ESPRIT on $\mathbf{Y}_{k,\Omega}$ to obtain $\hat{\phi}_{k}$.
\STATE Transform $\hat{\mathbf{Y}}_{k,cube}$ into $\mathbf{Y}_{k,\Psi}$, and then perform ESPRIT on $\mathbf{Y}_{k,\Psi}$ to obtain $\hat{\theta}_{k}$.
\STATE Transform $\hat{\mathbf{Y}}_{k,cube}$ into $\mathbf{Y}_{k,r}$, and then perform ESPRIT on $\mathbf{Y}_{k,r}$ to obtain $\hat{r}_{k}$.
		\FOR{$n_R^x=0$ to $N_R^x-1$}
		\FOR{$n_R^z=0$ to $N_R^z-1$}
		\STATE Extract $\mathbf{Y}_{k,v_{vir},n^x_R,n^z_R}$ from $\hat{\mathbf{Y}}_{k,cube}$, and then perform ESPRIT on $\mathbf{Y}_{k,v_{vir},n^x_R,n^z_R}$ to obtain $\hat{v}^{vir}_{k,n^x_R,n^z_R}$.
		\ENDFOR
		\ENDFOR
		\STATE Perform plane fitting on $\{(n_R^x,n_R^z,\hat{v}^{vir}_{k,n^x_R,n^z_R})|n_{R}^x =0,...,N^x_{R}\!-\!1; n_{R}^z =0,...,N^z_{R}\!-\!1\}$ to obtain $\hat{A}_{k}$, $\hat{B}_{k}$,
$\hat{C}_{k}$.
		\STATE Utilize $\hat{A}_{k}$, $\hat{B}_{k}$,
$\hat{C}_{k}$ to calculate and obtain $\hat{\omega}_{\phi,k},\hat{\omega}_{\theta,k},\hat{v}_{r,k}$.
	\end{algorithmic}
\end{algorithm}

Fig.~4 shows an example of virtual velocity estimation and fitting.  It can be seen from the figure that different antennas have observed different virtual velocities for the same dynamic target.  The 2D antenna index and the virtual velocity form a plane in the 3D coordinate system. Thus, we
can   recover the RV, HAV, and PAV of the target from these virtual velocities $\{(n_R^x,n_R^z,\hat{v}^{vir}_{k,n^x_R,n^z_R})|n_{R}^x =0,1,...,N^x_{R}-1; n_{R}^z =0,1,...,N^z_{R}-1\}$ based on equations (62) to (68).

\begin{figure*}[!t]
	\centering
	\subfloat[]{\includegraphics[width=60mm]{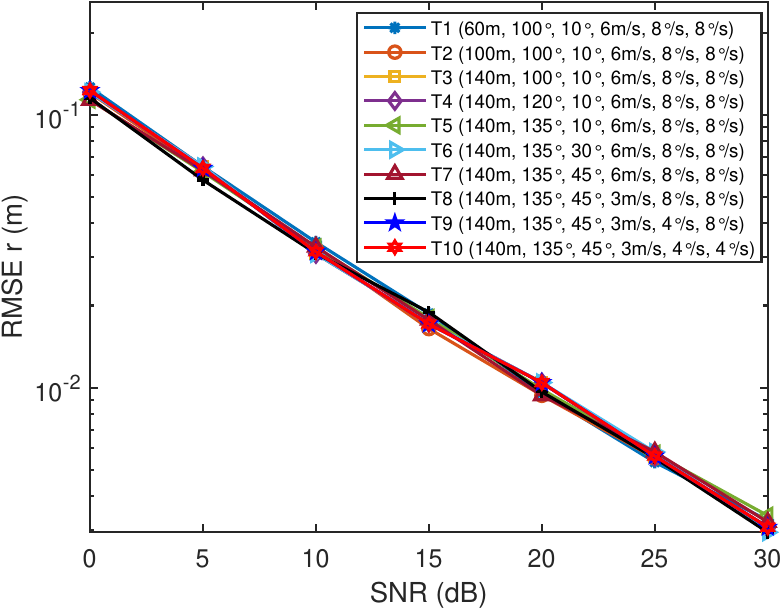}%
		\label{fig_first_case}}
	\hfil
	\subfloat[]{\includegraphics[width=60mm]{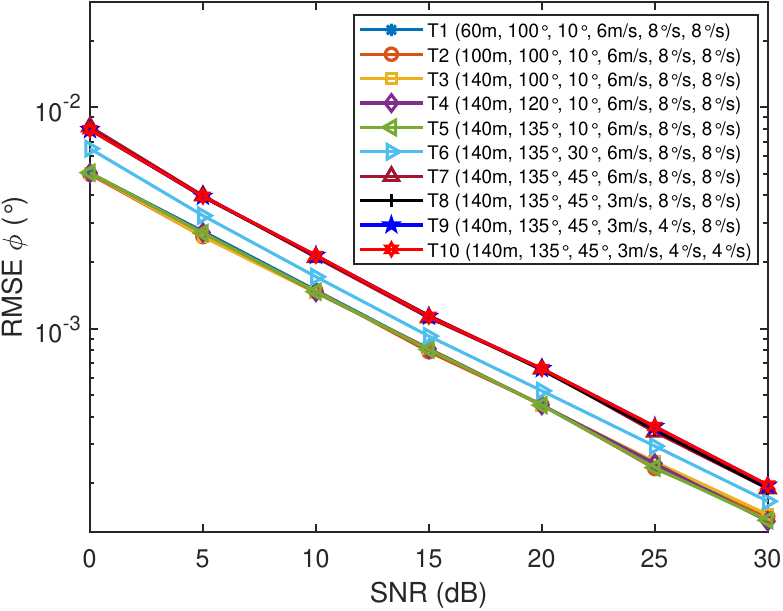}%
		\label{fig_first_case}}
	\hfil
	\subfloat[]{\includegraphics[width=60mm]{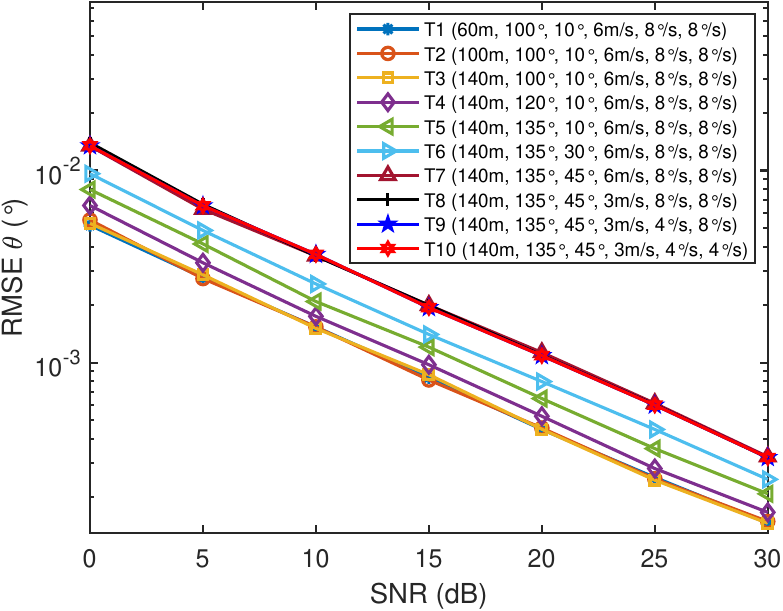}%
		\label{fig_first_case}}
	\hfil
	\subfloat[]{\includegraphics[width=60mm]{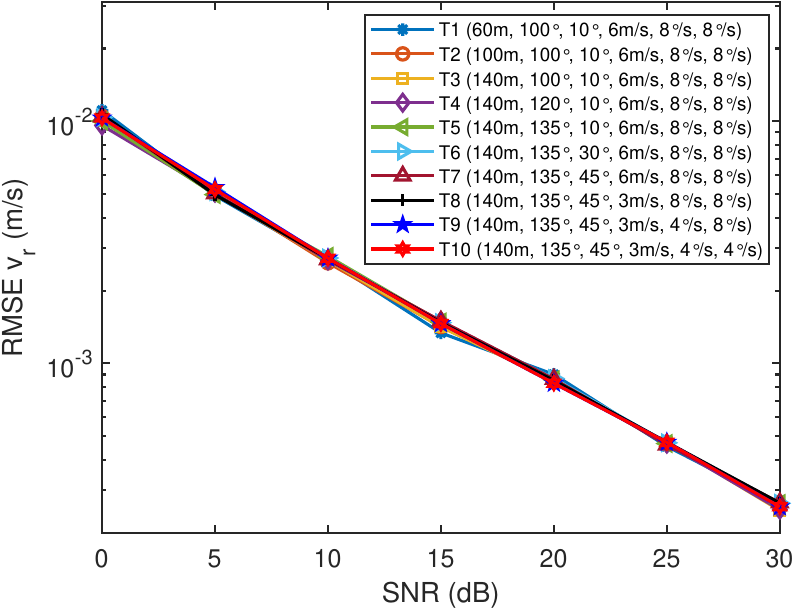}%
		\label{fig_first_case}}
	\hfil
	\subfloat[]{\includegraphics[width=60mm]{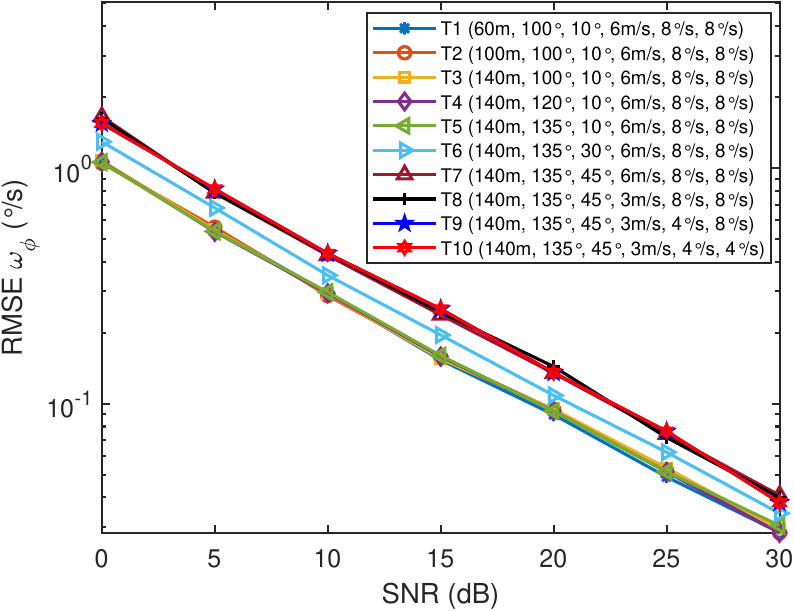}%
		\label{fig_first_case}}
	\hfil
	\subfloat[]{\includegraphics[width=60mm]{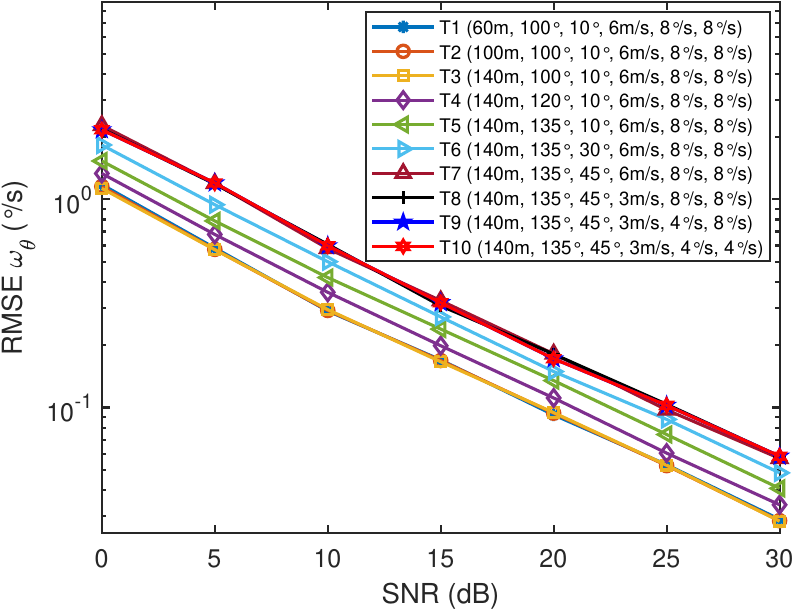}%
		\label{fig_second_case}}
	\caption{(a) The ${\rm RMSE}_r$ versus SNR curves of different targets.
(b) The ${\rm RMSE}_\phi$ versus SNR curves of different targets.
(c) The ${\rm RMSE}_\theta$ versus SNR curves of different targets. 
(d) The ${\rm RMSE}_{v_r}$ versus SNR curves of different targets.
(e) The ${\rm RMSE}_{\omega_\phi}$ versus SNR curves of different targets.
(f) The ${\rm RMSE}_{\omega_\theta}$ versus SNR curves of different targets. The legend represents the 6D parameters  $(r,\theta,\phi,v_r,\omega_\theta,\omega_\phi)$ of the target.}
	\label{fig_sim}
\end{figure*}

\section{Simulation Results}

In simulations, we set the lowest carrier frequency of  ISAC system as $f_0 = 28$ GHz, the subcarrier frequency interval as $\Delta f = 480$ kHz, and the antenna spacing as $d=\frac{1}{2}\lambda$. 
Specifically, to evaluate the performance of 6D motion parameters estimation, the root  mean square error (RMSE) of
Dis estimation,  HA estimation, PA estimation,  RV estimation, HAV estimation, and  PAV estimation  are defined as
${\rm RMSE}_r=\sqrt{\frac{\sum _{i=1}^{I}(\hat{r}_{s(i)}-r_{s})^2}{I}}$,
${\rm RMSE}_\theta=\sqrt{\frac{\sum _{i=1}^{I}(\hat{\theta}_{s(i)}-\theta_{s})^2}{I}}$,
${\rm RMSE}_\phi=\sqrt{\frac{\sum _{i=1}^{I}(\hat{\phi}_{s(i)}-\phi_{s})^2}{I}}$,   
${\rm RMSE}_{v_r}=\sqrt{\frac{\sum _{i=1}^{I}(\hat{v}_{r,s(i)}-v_{r,s})^2}{I}}$, 
${\rm RMSE}_{\omega_\theta}=\sqrt{\frac{\sum _{i=1}^{I}(\hat{\omega}_{\theta,s(i)}-\omega_{\theta,s})^2}{I}}$,
and ${\rm RMSE}_{\omega_\phi}=\sqrt{\frac{\sum _{i=1}^{I}(\hat{\omega}_{\phi,s(i)}-\omega_{\phi,s})^2}{I}}$, 
where $I$ is the number of the Monte Carlo runs,
$(r_{s},\theta_s,\phi_{s},v_{r,s},\omega_{\theta,s},\omega_{\phi,s})$ are the real parameters of the dynamic target, 
and $(\hat{r}_{s},\hat{\theta}_{s},\hat{\phi}_{s},\hat{v}_{r,s},\hat{\omega}_{\theta,s},\hat{\omega}_{\phi,s})$ are the estimated parameters of the target.

\begin{figure*}[!t]
\centering
\subfloat[]{\includegraphics[width=60mm]{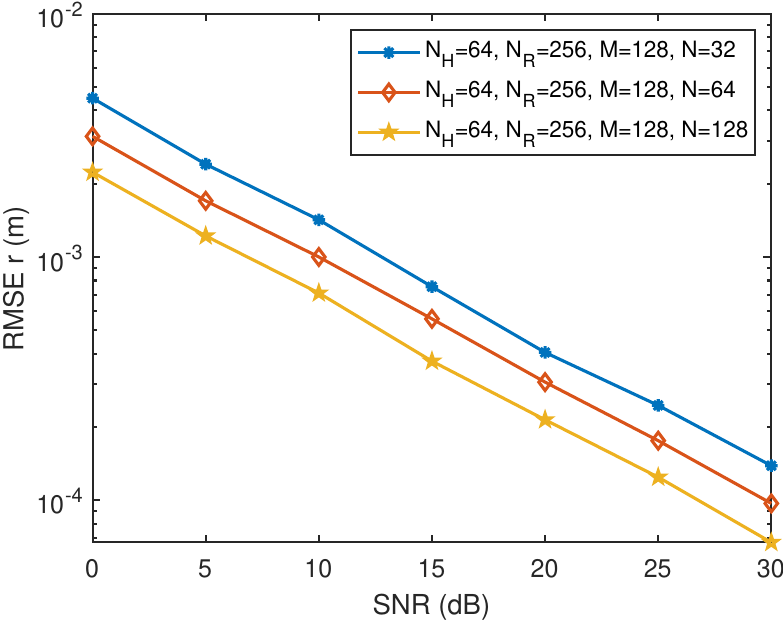}%
\label{fig_first_case}}
\hfil
\subfloat[]{\includegraphics[width=60mm]{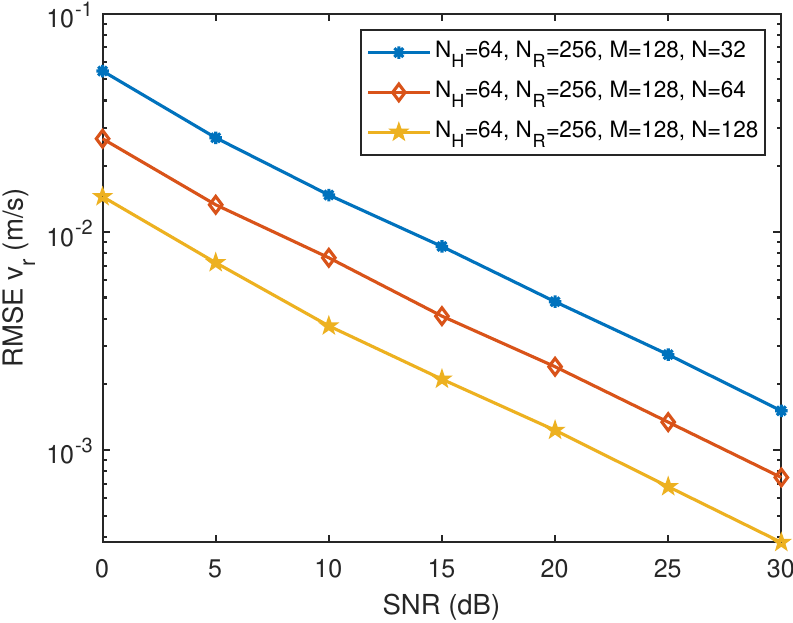}%
\label{fig_first_case}}
\hfil
\subfloat[]{\includegraphics[width=60mm]{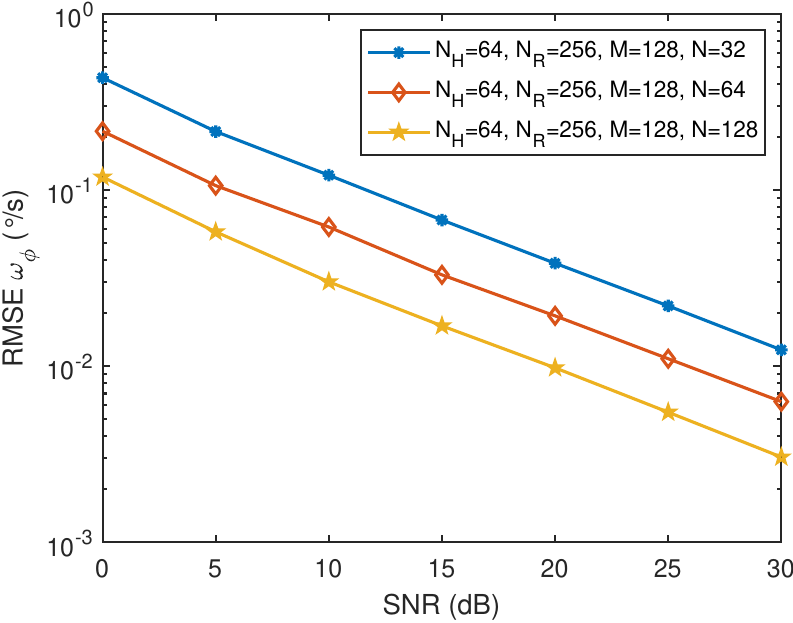}%
\label{fig_second_case}}
\caption{(a) Dis sensing performance under different numbers of OFDM symbols.
(b) RV sensing performance under different numbers of OFDM symbols.
(c) PAV sensing performance under different numbers of OFDM symbols.}
\label{fig_sim}
\end{figure*}

\begin{figure*}[!t]
\centering
\subfloat[]{\includegraphics[width=60mm]{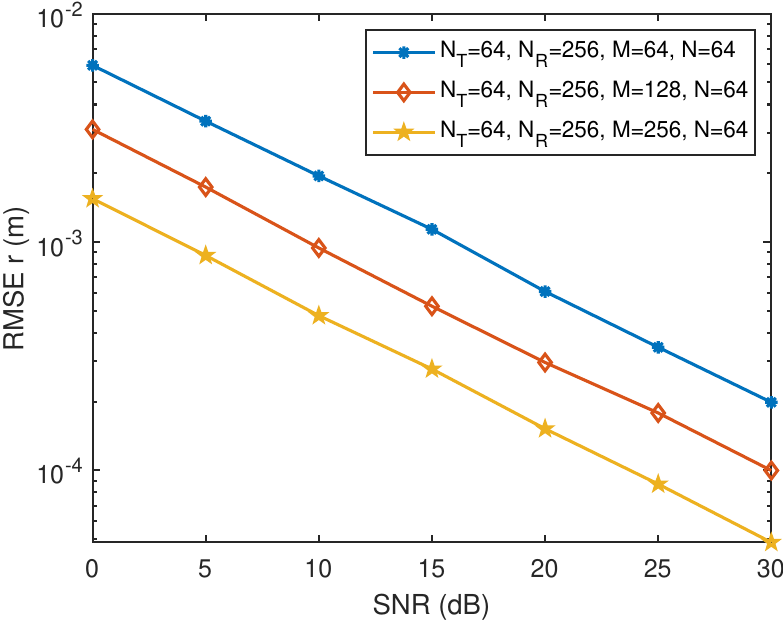}%
\label{fig_first_case}}
\hfil
\subfloat[]{\includegraphics[width=60mm]{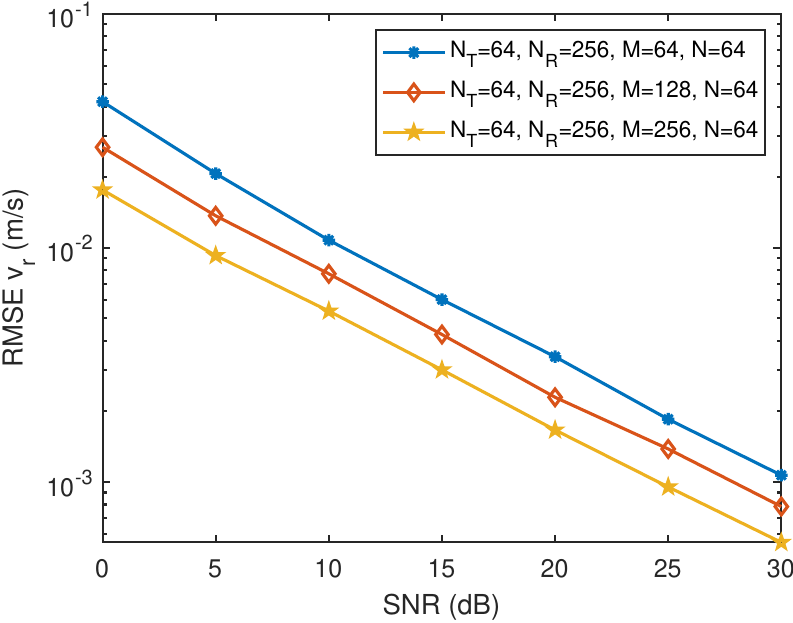}%
\label{fig_first_case}}
\hfil
\subfloat[]{\includegraphics[width=60mm]{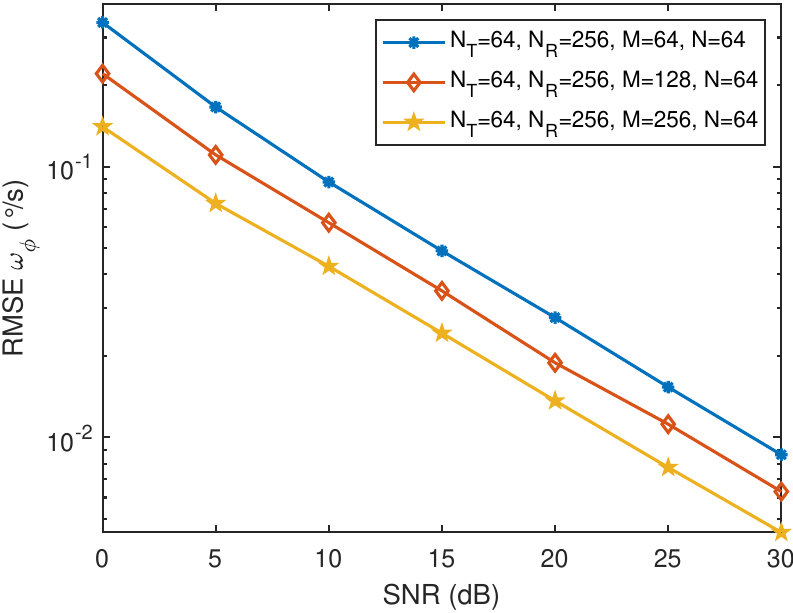}%
\label{fig_second_case}}
\caption{(a) Dis sensing performance under different numbers of subcarriers.
	(b) RV sensing performance under different numbers of subcarriers.
	(c) PAV sensing performance under different numbers of subcarriers.}
\label{fig_sim}
\end{figure*}

\subsection{The Performance of 6D Motion Parameters Estimation}

We set the number of subcarriers as $M=64$,
 the number of OFDM symbols as $N=64$,
the number of the antennas in HU-UPA as $N_H^x=64$ and $N_H^z=64$, and 
the number of the antennas in RU-UPA as $N_R^x=256$ and $N_R^z=256$.
Fig.~5 shows the  sensing RMSE of the proposed 6D motion parameters estimation scheme versus SNR for
different dynamic targets  with different motion parameters.
It can be seen from the figure that the 
${\rm RMSE}_{r}$, ${\rm RMSE}_{\theta}$, ${\rm RMSE}_{\phi}$,
${\rm RMSE}_{v_r}$,  ${\rm RMSE}_{\omega_{\theta}}$ and 
${\rm RMSE}_{\omega_{\phi}}$ gradually decrease with the increase of SNR. 
When SNR $=10$ dB, the average estimating RMSEs are
${\rm RMSE}_{r}=0.0319m$, 
${\rm RMSE}_{\theta}=0.0025^\circ$, 
${\rm RMSE}_{\phi}=0.0017^\circ$,
${\rm RMSE}_{v_r}=0.0027m/s$,  
${\rm RMSE}_{\omega_{\theta}}=0.4562^\circ/s$ and 
${\rm RMSE}_{\omega_{\phi}}=0.3552^\circ/s$.
When SNR increases to $20$ dB, the average estimating RMSEs decrease to
${\rm RMSE}_{r}=0.0099m$, 
${\rm RMSE}_{\theta}=0.0008^\circ$, 
${\rm RMSE}_{\phi}=0.0006^\circ$,
${\rm RMSE}_{v_r}=0.0009m/s$,  
${\rm RMSE}_{\omega_{\theta}}=0.1382^\circ/s$ and 
${\rm RMSE}_{\omega_{\phi}}=0.1146^\circ/s$.
Different from most existing ISAC studies\cite{9529026,10048770,8827589,2023arXiv230715074J,10403776,9947033}, where only the RV of far-field dynamic target can be estimated based on a single BS, Fig.~5  indicates  that the proposed 6D motion parameters estimation scheme confirms that one single BS can effectively estimate the HAV and PAV of  dynamic target.

Besides, it is found from Fig.~5(a) and  Fig.~5(d) that under the same system parameter settings, the Dis estimation and RV estimation performance with different motion parameters are basically consistent. 
However, it is seen from Fig.~5(b) and  Fig.~5(e) that under the same system parameter settings,   the accuracy of PA estimation and PAV estimation gradually improves as the target approaches $\phi = 0^\circ$, mainly because the MIMO array has narrower beamwidth near $0^\circ$, which thus improves the accuracy of PA estimation. Since the PAV estimation depends on the PA change of the target, the  narrower beam near $0^\circ$ also brings higher PAV estimation accuracy.
Noting that $\theta=90^\circ$ is located at the center of the MIMO array, we see from Fig.~5(c) and  Fig.~5(f) that when $\phi$ is fixed,  the accuracy of HA estimation and HAV estimation gradually improves as the target approaches $\theta = 90^\circ$.  Furthermore, according to Eq. (55) and Eq. (67), the estimating error of $\phi$ and $\omega_{\phi}$ further affects the estimation accuracy of $\theta$ and $\omega_{\theta}$.

\subsection{Impact of System Parameters on Estimation Performance}

\begin{figure*}[!t]
\centering
\subfloat[]{\includegraphics[width=60mm]{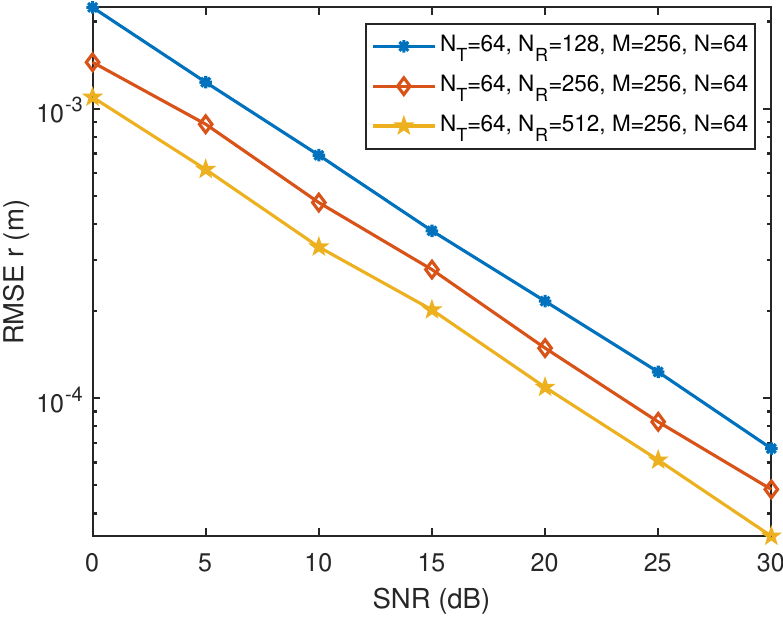}%
\label{fig_first_case}}
\hfil
\subfloat[]{\includegraphics[width=60mm]{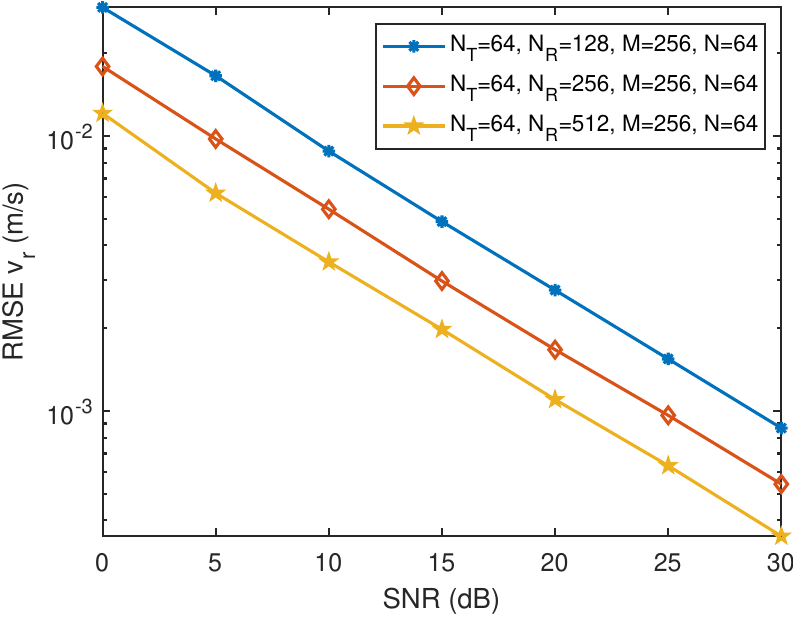}%
\label{fig_first_case}}
\hfil
\subfloat[]{\includegraphics[width=60mm]{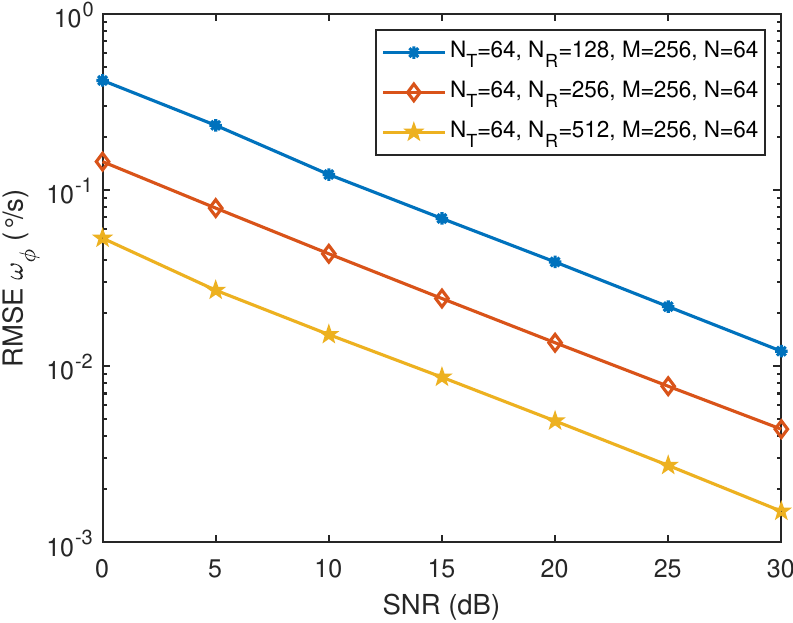}%
\label{fig_second_case}}
\caption{(a) Dis sensing performance under different numbers of receiving antennas.
	(b) RV sensing performance under different numbers of receiving antennas.
	(c) PAV sensing performance under different numbers of receiving antennas.}
\label{fig_sim}
\end{figure*}

We set the number of subcarriers as $M=128$,
 the number of OFDM symbols as $N=64$,
the number of the antennas in HU-UPA as $N_H^x=1$ and $N_H^z=64$, and 
the number of the antennas in RU-UPA as $N_R^x=1$ and $N_R^z=256$. 
We take the estimation of  the  dynamic target with motion parameters ($r=120m$, $\theta=90^\circ$, $\phi=20^\circ$,
 $v_r=15m/s$, $\omega_{\theta}=0^\circ/s$, $\omega_{\phi}=8^\circ/s$) as an example, and
  investigate the impact of system parameter settings on the performance of 6D motion parameters estimation.

Fig.~6 shows the variation curves of Dis estimation, RV estimation, and PAV estimation versus SNR under different number of  OFDM symbols. It can be seen from Fig.~6(a)  that the ${\rm RMSE}_r$ gradually decreases as the number of OFDM symbols $N$ increases. This is because more OFDM symbols bring more observations to the distance array, making the estimation of the covariance matrix of the distance array more accurate, and  thereby improving the accuracy of Dis estimation.
Besides,  it can be found from Fig.~6(b) and Fig.~6(c) that the ${\rm RMSE}_{v_r}$ and the ${\rm RMSE}_{\omega_\phi}$ significantly decrease with the increase of $N$. This is because more OFDM symbols form a larger virtual velocity array, making the estimation of RV and PAV more accurate.

Fig.~7 shows the variation curves of  estimating RMSEs versus SNR under different number of  subcarriers. 
 It can be seen from Fig.~7(a) that the ${\rm RMSE}_r$ gradually decreases with the increase of the number of subcarriers $M$, because more subcarriers can form a larger distance array, thereby improving the accuracy of Dis estimation. It can be found from  Fig.~7(b) and Fig.~7(c) that the  ${\rm RMSE}_{v_r}$ and the ${\rm RMSE}_{\omega_\phi}$  gradually decrease with the increase of $M$. This is because more subcarriers bring more observations to the virtual velocity array, making the covariance matrix estimation of the virtual velocity array more accurate, thereby improving the estimation accuracy of RV and PAV.

Fig.~8 shows the variation curves of  estimating RMSEs versus SNR under different number of antennas.
 It is seen from Fig.~8(a) that the ${\rm RMSE}_r$ gradually decreases as the number of antennas $N_R$ increases. This is because more receiving antennas provide more observations for  distance array, thereby improving the accuracy of Dis estimation.
More importantly, it can be observed from  Fig.~8(b) and Fig.~8(c) that the  ${\rm RMSE}_{v_r}$ and the ${\rm RMSE}_{\omega_\phi}$ gradually decrease with the increase of $N_R$.
This is because when there are more receiving antennas measuring the virtual velocity, the system can better fit the virtual velocity plane, thereby more accurately recovering the RV and PAV of dynamic target.

\section{Conclusions}

In this paper, we proposed a novel scheme to estimate the 6D motion parameters of dynamic target  for monostatic ISAC system.
Specifically, we provided a generic ISAC framework for dynamic target sensing based on massive MIMO array. 
Next, we derived the relationship between the sensing  channel of ISAC BS and the 6D motion parameters of dynamic target, based on which  we employed the  array signal processing methods to estimate the dynamic target's  HA, PA, Dis, and \emph{virtual velocity}. 
Then we adopted plane  fitting to estimate the RV, HAV, and PAV of  dynamic target from these virtual velocities. 
Simulation results  demonstrated the effectiveness of the proposed 6D motion parameters estimation scheme,  confirming a new finding  that one single BS with massive MIMO array is capable of estimating the HAV and PAV  of dynamic target.

\bibliographystyle{ieeetr}
\bibliography{paper9.bib}

\vfill

\end{document}